\documentclass[conference]{IEEEtran}

\usepackage[utf8]{inputenc}
\usepackage[english]{babel}
\usepackage{cite}
\usepackage{amsmath,amssymb,amsfonts}
\usepackage{subcaption}
\usepackage{graphicx}
\usepackage{textcomp}
\usepackage{xcolor}
\usepackage{mathrsfs}
\usepackage[utf8]{inputenc}
\usepackage{comment}
\usepackage[ruled,vlined]{algorithm2e}
\usepackage{mathtools}
\usepackage[font=small,labelfont=bf]{caption}
\usepackage{blkarray, bigstrut}
\usepackage{cases}
\usepackage{amsmath}
\usepackage{orcidlink}

\usepackage[utf8]{inputenc}

\title{Network-Controlled  Physical-Layer Security: Enhancing Secrecy through Friendly Jamming}

\author{\IEEEauthorblockN{Sayed Amir Hoseini$^{\orcidlink{0000-0002-3105-4218}}$, Parastoo Sadeghi$^{\orcidlink{0000-0002-9965-9483}}$, Faycal Bouhafs$^{\orcidlink{0000-0001-6626-7881}}$, Neda Aboutorab$^{\orcidlink{0000-0001-9118-3866}}$ and Frank den Hartog$^{\orcidlink{0000-0001-5293-6140}}$}
\IEEEauthorblockA{School of Engineering and IT, University of New South Wales, Canberra, Australia \\
Email: \{s.a.hoseini, p.sadeghi, f.bouhafs, n.aboutorab, frank.den.hartog\}@unsw.edu.au}}

\newcommand\numberthis{\addtocounter{equation}{1}\tag{\theequation}}
\usepackage[pscoord]{eso-pic}
\newcommand{\placetextbox}[3]{
\setbox0=\hbox{#3}
\AddToShipoutPictureFG*{
\put(\LenToUnit{#1\paperwidth},\LenToUnit{#2\paperheight}){\vtop{{\null}\makebox[0pt][c]{#3}}}
}
}
\placetextbox{.21}{0.055}{\small{978-1-6654-9792-3/22/\$31.00~\copyright2022 IEEE}}
\begin{document}
\bstctlcite{IEEEexample:BSTcontrol}
\maketitle

\begin{abstract}
The broadcasting nature of the wireless medium makes exposure to eavesdroppers a potential threat. Physical Layer Security (PLS) has been widely recognized as a promising security measure complementary to encryption. It has recently been demonstrated that PLS can be implemented using off-the-shelf equipment by spectrum-programming enhanced Software-Defined Networking (SDN), where a network controller is able to execute intelligent access point (AP) selection algorithms such that PLS can be achieved and secrecy capacity optimized. In this paper we provide a basic system model for such implementations. We also introduce a novel secrecy capacity optimization algorithm, in which we combine intelligent AP selection with the addition of Friendly Jamming (FJ) by the not-selected AP.
\end{abstract}

\begin{IEEEkeywords}
Artificial Noise, Secrecy, Physical-Layer Security, SDN, Programmable Networks, Friendly Jamming
\end{IEEEkeywords}
\section{Introduction}
Society has become unthinkable without wireless devices. We have become reliant on wireless communication technologies to exchange personal and sometimes confidential data. The broadcasting nature of the wireless medium makes exposure to eavesdroppers a potential threat. So far, this threat has mostly been mitigated by encrypting the wireless link and the information transmitted. Such a solution assumes that eavesdroppers lack the computational resources and knowledge of the network parameters to break the encryption. While this assumption still hold for many scenarios today, eavesdroppers' capabilities are rapidly improving.

Physical Layer Security (PLS) has been widely recognized as a promising complementary security measure. PLS limits the information that can be intercepted at the bit-level by making it impossible for an eavesdropper to decode any data at the physical layer \cite{wyner1975wire}. If executed well, PLS can thus provide perfect secrecy. Until recently, implementing PLS in a practical and cost-effective way was a challenge \cite{ryland2018software}. Most techniques proposed in the literature involve major signal-processing efforts to scramble the communication channel effectively for the eavesdropper while simultaneously optimizing the throughput for the legitimate station. 

In recent work \cite{hoseini2022ccnc}, we demonstrated that PLS can be realized using off-the-shelf equipment by tackling the problem at the network-level. The idea is that a wireless network typically contains not just one wireless access point (AP), but many APs to which a legitimate station could possibly connect. Using a relatively new enhancement of Software-Defined Networking (SDN) specifically for wireless networks, called spectrum programming \cite{bouhafs2018wi}, it is now possible to execute intelligent AP selection algorithms in a way that is completely transparent to the connecting station. We investigated two such algorithms in earlier work. In \cite{hoseini2022ccnc}, we had the legitimate station always connect to the AP that is least beneficial to the eavesdropper, and in \cite{Faycal2020globecom} the AP was selected that maximized the secrecy capacity for the legitimate station. The secrecy capacity is the maximum capacity a legitimate station can achieve under the condition of full secrecy while connected to a given AP.

In this paper, after discussing related work in section II, we provide a basic system model of the proof-of-concept described in \cite{hoseini2022ccnc}. The model enables us to introduce a novel secrecy capacity optimization algorithm, described in section IV, in which we combine intelligent AP selection based on maximizing secrecy capacity \cite{Faycal2020globecom} with the addition of Friendly Jamming (FJ) by the not-selected AP. In section V we show that providing such a FJ signal to the eavesdroppers significantly improves secrecy in the network beyond what can be achieved with intelligent AP selection. Although the concepts of AP selection and FJ have been suggested earlier in isolation, this is the first time they are combined and supported by a single robust theoretical framework, and in section VI we discuss its applicability in realistic networks.

%we may want to discuss Globecom and CCNC approaches in AP selection here or in other sections

\section{Related Work}
Techniques proposed in the literature to achieve PLS can be categorized as Channel Coding techniques, Channel Control techniques, Power Control techniques, and Artificial Noise (AN) techniques. Channel coding techniques introduce robust coding schemes and randomization in the transmitted signal to make it difficult for eavesdroppers to decode the intercepted signal \cite{harrison2013coding,harrison2009physical}. Channel Control focuses on manipulating the radio channel parameter and monitoring the channel to detect the presence of eavesdroppers \cite{sperandio2002wireless,li2005mimo}. Power Control techniques try to achieve secrecy by controlling the power and direction of the signal transmitted to increase the capacity at the legitimate station and degrade the capacity at the eavesdropper, for instance by using multiple antennas \cite{chen2016survey}. AN techniques, which could also be considered as power control techniques, aim to generate jamming signals to achieve PLS especially in situations where the eavesdropper is closer to the source than the legitimate station. AN-based PLS has been investigated extensively as an approach to achieve PLS for wireless communications. Early work in this area focused on providing AN-based PLS with partial or no knowledge of the Channel State Information (CSI) at the legitimate and eavesdropping stations \cite{liu2013practical,lin2013secrecy,tsai2014power,zhang2013design}. 

These techniques have typically approached the problem at the link-level, focusing  on  the  individual  wireless  connections  between sender and receiver and, so far, have proven to be very hard, if not impossible, to implement. Our earlier work, as presented in \cite{Faycal2020globecom} and \cite{hoseini2022ccnc}, shows a solution by approaching the problem at the network-level using multiple APs and intelligent AP selection. The solution takes advantage of SDN-based spectrum programming as presented in \cite{bouhafs2018wi}, where an enhanced programmable controller has full, up-to-date knowledge of the CSI across the network, and has fine-grained control over the radio parameters of each AP. 

In this paper, we show that secrecy in the network can be improved by providing a FJ signal to the eavesdroppers. The idea of FJ in itself is not new. Work presented in \cite{jorgensen2007shout} and \cite{vilela2011wireless} propose an architecture to help realize PLS where two APs, namely AP1 and AP2, are deployed with the legitimate station transmitting data to AP1. During this transmission, AP2 sends a pre-determined signal to AP1, which will act as a jamming signal to any eavesdropping station in the vicinity of the communication. More recent contributions proposing the orchestration of system-level interference to achieve PLS use new techniques such as intelligent reflecting surface \cite{hong2020artificial} and non-orthogonal multiple access \cite{gong2021enhancing}. These contributions are all limited to the theoretical domain and do not offer a practical approach to implement the proposed solutions. The novelty of our work lies in enhancing secrecy by \textit{combining} intelligent AP selection with orchestrated network-controlled generation of FJ in a way that suits today's hardware and software, supported by a robust theoretical framework.  

\section{System Model}\label{sec:model}
\begin{figure}
    \centering
    \includegraphics[width=0.45\textwidth]{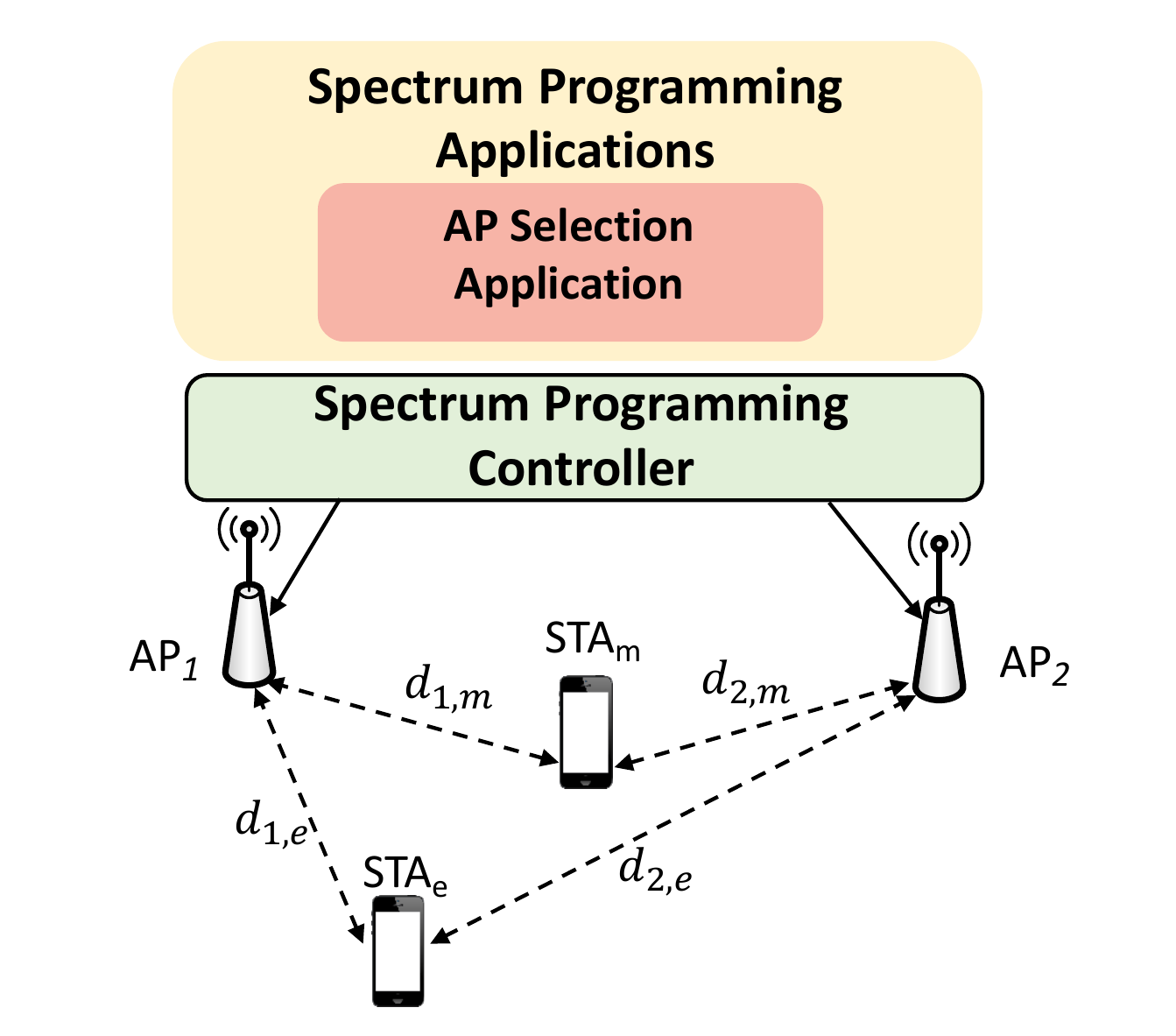}
    \caption{Network-level PLS: the legitimate station STA$_{m}$ is connected to the AP that can provide the best secrecy capacity in the presence of an eavesdropper STA$_e$.}
    \label{fig:system}
\end{figure}

We here consider a wireless system model as shown in \figurename~\ref{fig:system},  where a legitimate station is trying to connect to, and receive information from, a wireless network in the presence of an eavesdropping station. The remainder of the paper assumes the network to be based on 2.4 GHz Wi-Fi, but the  concept  can  easily  be  generalized  to  other  wireless networks.  

It is known that PLS can, in theory, be achieved when the Shannon capacity of the legitimate station is higher than the Shannon capacity of the eavesdropping station (under a range of conditions as laid out in \cite{wyner1975wire}). Without loss of generality, we assume downstream traffic, i.e., from one of the Wi-Fi APs to the legitimate station. This is reasonable for situations where confidential information is provided by servers in the network and is only offered for consumption to legitimate clients. 

% \subsection{Access Point Selection (Wi-5)}\label{sec:AP selection}
Consider a scenario where there are two APs (denoted by AP$_1$ and AP$_2$), one legitimate station (denoted by STA$_m$) and an eavesdropper (denoted by STA$_e$). We employ the AP selection mechanism proposed in \cite{Faycal2020globecom} that exploits the principles of PLS to assign the AP that can provide the highest secrecy capacity for STA$_m$. To better explain the AP selection mechanism, a summary of the notation used in this paper is presented in Table \ref{tab:notations}. We assume that the location of the APs, STA$_m$ and STA$_e$ are known. The latter is of course hard when the  eavesdropper is passive, but various proposals have been made in the literature to overcome this problem (see, e.g., \cite{chaman2018ghostbuster}). We will discuss this further in Section VI.  For simplicity of exposition, we will model the channel between the APs and both stations using a path loss model \cite{goldsmith2005wireless}. Our optimization method is, however, extendable to the case where block-fading CSI is available at the spectrum programming controller.  

\newcommand{\opertype}[1]{\begin{minipage}{65mm}\centering #1\end{minipage}}
\begin{table}[htb]
    \caption{Summary of notation and units}
    \centering
    \begin{tabular}{|lp{6.8cm	}|lp{8cm}|p{30cm}}
    \hline
         AP$_n$& Either one of the Access Points ($n$ is either 1 or 2)\\
         \hline
         STA$_m$&  Legitimate station\\
         \hline
         STA$_e$&  Eavesdropper station\\
         \hline
         $W$& Channel bandwidth, Hz\\
         \hline
         $f_0$& Operating center frequency, Hz\\
         \hline
         $C$& Speed of light, $\sim 3\times 10^8$ m/s\\
         \hline
         $P_{t_n}$& Transmit power of AP$_n$, Watt\\
         \hline
         $P_{t_n}^{\max}$& Maximum transmit power of AP$_n$, Watt\\
         \hline
         $d_{n,m}$& Distance between STA$_m$ and AP$_n$, m\\
         \hline
         $d_{n,e}$& Distance between STA$_e$ and AP$_n$, m\\
         \hline
         $\alpha$& Path loss exponent (typical values are:\\
         & $\alpha=2$ for free space,\\
         &$\alpha=2.7\sim 3.5$ for urban area,\\
         &$\alpha=1.6\sim 1.8$ for indoor (line-of-sight))\\
        \hline
          $P_{0,n}$& {Free-space received power from AP$_n$ at reference distance $d_0$: $P_{0,n} = P_{t_n}(\frac{C}{4\pi f_0d_0})^2$, Watt}\\
       \hline
          $P_{n}$& {Distance-corrected  power used in capacity formulas: $P_n = P_{0,n}d_0^{\alpha}$,  Watt $m^\alpha$}\\
         \hline
         SINR$_{n,m}$& SINR at station STA$_m$ when connected to AP$_n$\\     
         \hline
         SINR$_{n,e}$& SINR at station STA$_e$ when connected to AP$_n$\\     
         \hline
         $I_{n,m}$& Interference experienced at STA$_m$, Watt\\
         \hline
         $I_{n,e}$& Interference experienced at STA$_e$, Watt\\  
         \hline
         $N_m$& Noise experienced by STA$_m$, Watt\\
         \hline
         $N_e$& Noise experienced by STA$_e$, Watt\\
         \hline
         $ C_{n,m}$& Channel capacity  between AP$_n$ and STA$_m$, bits/s\\
         \hline
         $ C_{n,e}$& Channel capacity  between AP$_n$ and STA$_e$, bits/s\\
         \hline
    \end{tabular}
    \label{tab:notations}
\end{table}

Let us assume that AP$_n$ ($n$ is either 1  or 2) is the considered candidate for downlink data transmission. The received power at STA$_m$ and STA$_e$ from $AP_n$ is $P_{n}d_{n,m}^{-\alpha}$ and $P_{n}d_{n,e}^{-\alpha}$, respectively. Therefore, the Shannon capacity of the channel between AP$_n$ and the legitimate station STA$_m$  is given as 
\begin{equation}\label{eq:legit_cap}
    C_{n,m}=W\log(1+\text{SINR}_{n,m}) = W\log\left(1+\frac{P_nd_{n,m}^{-\alpha}}{I_{n,m}+N_{m}}\right).
\end{equation}
with SINR$_{n,m}$ the Signal to Interference plus Noise Ratio at STA$_{m}$ from AP$_{n}$. Similarly, the Shannon capacity of the channel between AP$_n$ and the eavesdropper STA$_e$ is
\begin{equation}\label{eq:eve_cap}
    C_{n,e}=W\log(1+\text{SINR}_{n,e})=W\log\left(1+\frac{P_nd_{n,e}^{-\alpha}}{I_{n,e}+N_{e}}\right),
\end{equation}
where all $\log$s are in base 2 and, therefore, capacities are measured in bits/s. The terms $I_{n,m}$ and $I_{n,e}$  measure the interference experienced at STA${_m}$ and STA$_e$, respectively, as further elaborated in Section IV.

Based on the principles of PLS, STA$_m$ can securely communicate with AP$_n$ if $C_{n,m}>C_{n,e}$. The proposed AP selection mechanism in \cite{Faycal2020globecom} then connects STA$_m$ to the AP that provides the maximum secrecy capacity (i.e., maximum $C_{n,m}-C_{n,e}$ value among AP choices $n = 1$ or $n=2$).  Thus, the secrecy capacity is maximized through finding the solution
\begin{equation}\label{eq:maxsecrecy}
    i=\arg\max_{n \in \{1,2\}}  ({C_{n,m}-C_{n,e}}).
\end{equation}
Therefore, AP$_i$ is the selected AP for transmission of information to STA$_m$.  The other access point AP$_j$, $j \neq i$ is ``idle", as far as data traffic is concerned. 

\section{Proposed Friendly Jamming}
We here propose to have the not-selected AP in \eqref{eq:maxsecrecy} (which may also call idle AP) generating an optimal Friendly Jamming (FJ) signal, in addition to employing the AP selection mechanism described in Section \ref{sec:model}. The idle AP$_j$ tries to jam STA$_e$ to reduce its SINR, which in turns reduces the channel capacity of the eavesdropper ($C_{e,n}$) and, therefore, improves the secrecy capacity of STA$_m$.  For ease of exposition, we assume that STA$_e$ and STA$_m$ have the same noise powers for their receivers, i.e. $N = N_m=N_e$. We also assume that at the time of AP selection based on the mechanism in \eqref{eq:maxsecrecy}, the \emph{ambient} interference (without FJ generation) was zero. In our problem formulation and subsequent performance evaluation, this will allow us to evaluate the improvement from optimal FJ generation compared to a baseline system (which is free from ambient/extra interference) in a meaningful way. However, the method presented below can be extended to cater for the general case. 

Given the above, the interference experienced by STA$_m$ and STA$_e$ from the FJ generating AP$_j$ is $P_jd_{j,m}^{-\alpha}$ and $P_jd_{j,e}^{-\alpha}$, respectively. Therefore, we specify \eqref{eq:legit_cap} and \eqref{eq:eve_cap} as

\begin{equation}
    C_{i,m}=W\log\left(1+\frac{P_id_{i,m}^{-\alpha}}{P_jd_{j,m}^{-\alpha}+N}\right),
\end{equation}
\begin{equation}
    C_{i,e}=W\log\left(1+\frac{P_id_{i,e}^{-\alpha}}{P_jd_{j,e}^{-\alpha}+N}\right).
    \label{eq:eve_cap2}
\end{equation}

The goal here is to find the optimal interference power transmitted from AP$_j$, i.e. the optimal $P_j$, to maximize the secrecy capacity of the downlink transmission from AP$_i$ to STA$_m$. In order to achieve that, we fix the power for the main data AP$_i$, $P_i$, and find the optimal $P_j$ that maximizes $C_{i,m}-C_{i,e}$ defined as follows
\begin{equation}\label{eq:log}
\begin{split}
       C_{i,m}-C_{i,e}&=W\log(1+\frac{P_id_{i,m}^{-\alpha}}{P_jd_{j,m}^{-\alpha}+N})\\
       &-W\log(1+\frac{P_id_{i,e}^{-\alpha}}{P_jd_{j,e}^{-\alpha}+N})\\
    &=W\log(\frac{P_jd_{j,m}^{-\alpha}+N+P_id_{i,m}^{-\alpha}}{P_jd_{j,m}^{-\alpha}+N})\\
    &-W\log(\frac{P_jd_{j,e}^{-\alpha}+N+P_id_{i,e}^{-\alpha}}{P_jd_{j,e}^{-\alpha}+N})\\
    &=W\log(\frac{P_jd_{i,m}^{\alpha}+Nd_{i,m}^{\alpha}d_{j,m}^{\alpha}+P_id_{j,m}^{\alpha}}{P_jd_{i,m}^{\alpha}+Nd_{i,m}^{\alpha}d_{j,m}^{\alpha}})\\
    &-W\log(\frac{P_jd_{i,e}^{\alpha}+Nd_{i,e}^{\alpha}d_{j,e}^{\alpha}+P_id_{j,e}^{\alpha}}{P_jd_{i,e}^{\alpha}+Nd_{i,e}^{\alpha}d_{j,e}^{\alpha}})\\
    &=W\log(\frac{P_jd_{i,m}^{\alpha}+Nd_{i,m}^{\alpha}d_{j,m}^{\alpha}+P_id_{j,m}^{\alpha}}{P_jd_{i,m}^{\alpha}+Nd_{i,m}^{\alpha}d_{j,m}^{\alpha}}\\
    &\times \frac{P_jd_{i,e}^{\alpha}+Nd_{i,e}^{\alpha}d_{j,e}^{\alpha}}{P_jd_{i,e}^{\alpha}+Nd_{i,e}^{\alpha}d_{j,e}^{\alpha}+P_id_{j,e}^{\alpha}}).
\end{split}
\end{equation}
To simplify, let us refer to the argument inside the last logarithmic term as $f(P_i, P_j)$. Therefore, $C_{i,m}-C_{i,e}$ can simply be expressed as
\begin{equation}\label{eq:log_opt}
C_{i,m}-C_{i,e}=W\log(f(P_i,P_j)),
\end{equation}
where $f(P_i,P_j)$ is given by
\begin{equation}\label{eq:f}
f(P_i,P_j)=\frac{{P_j}^2A+P_jB+P_iP_jC+P_iD+K}{{P_j}^2A+P_jB+P_iP_jE+P_iF+K},
\end{equation}
and $A, B, C, D, E, F$ and $K$ are defined as follows:
\begin{equation}
\begin{split}
A&=d_{i,m}^{\alpha}d_{i,e}^{\alpha},\\
B&=Nd_{i,e}^{\alpha}d_{j,e}^{\alpha}d_{i,m}^{\alpha}+Nd_{i,m}^{\alpha}d_{j,m}^{\alpha}d_{i,e}^{\alpha},\\
C&=d_{j,m}^{\alpha}d_{i,e}^{\alpha},\\
D&=Nd_{i,e}^{\alpha}d_{j,e}^{\alpha}d_{j,m}^{\alpha},\\
E&=d_{i,m}^{\alpha}d_{j,e}^{\alpha},\\
F&=Nd_{i,m}^{\alpha}d_{j,m}^{\alpha}d_{j,e}^{\alpha},\\
K&=N^2d_{i,m}^{\alpha}d_{j,m}^{\alpha}d_{i,e}^{\alpha}d_{j,e}^{\alpha}.\\
\end{split}
\end{equation}
The partial derivative of $f$ with respect to $P_j$ is
\begin{equation}
\begin{split}\label{eq:derivative}
\frac{\partial f}{\partial P_j}=\frac{P_j^2a+P_jb+c}{{P_j}^2A+P_jB+P_iP_jE+P_iF+K},
\end{split}
\end{equation}
where $a, b$ and $c$ are defined as
\begin{equation*}
\begin{split}
a&=2P_iAE+P_iAC-2P_iAC-P_iAE,\\
b&=2P_iAF-2P_iAD,\\
c&=P_iBF+P_i^2FC+P_iKC-P_iBD-P_i^2ED-P_iKE.
\end{split}
\end{equation*}
Let $P^{\max} := P_{t_{j}}^{\max}\left(\frac{C}{4\pi f_0 d_0}\right)^2 d_0^\alpha$ and denote the  two quadratic solutions of $\frac{\partial f}{\partial P_j} = 0$ by $Q_j^{1,2}$. Note that $Q_j^{1,2}$ may be negative or go above $P^{\max}$.
Therefore, we need to adjust the above roots according to the physical system constraints:
\begin{equation}
    P_j^{k}= \min\left\{\max\{Q_j^{k},0\}, P^{\max}\right\}, \quad k = 1,2.
\end{equation} 
In addition, two boundary power candidates $P_j^{3} = 0$ (no FJ at all) or $P_j^{4} = P^{\max}$ (max FJ power) also need to be considered. In summary, the optimal FJ power solution is the one among the four candidates that gives the best secrecy capacity $C_{i,m}-C_{i,e}$. That is,
\begin{equation}\label{eq:optimalAN}
    P_j^{\text{Optimal}}= \arg\min_{k \in \{1,\cdots, 4\}}W\log(f(P_i,P_j^{k})).
\end{equation}

\section{Performance Evaluation}
In this section, we evaluate the performance of the proposed algorithm in MatLab. We simulated three Wi-Fi systems: a) a normal Wi-Fi system where STA$_m$ is associated to the AP with the highest SINR regardless of the eavesdropper's location; b) the smart AP system based on \cite{Faycal2020globecom} where STA$_m$ is associated to AP$_i$ that provides the highest secrecy capacity according to \eqref{eq:maxsecrecy}; and c) the enhanced smart AP where the idle access point AP$_j$ generates FJ to increase the secrecy of communication between STA$_m$ and AP$_i$ according to \eqref{eq:optimalAN}.

\begin{figure*}
     \centering
     \begin{subfigure}[b]{0.297\textwidth}
         \hspace{-4.7mm}
         \includegraphics[trim=0 0 0 0, clip, width=\textwidth]{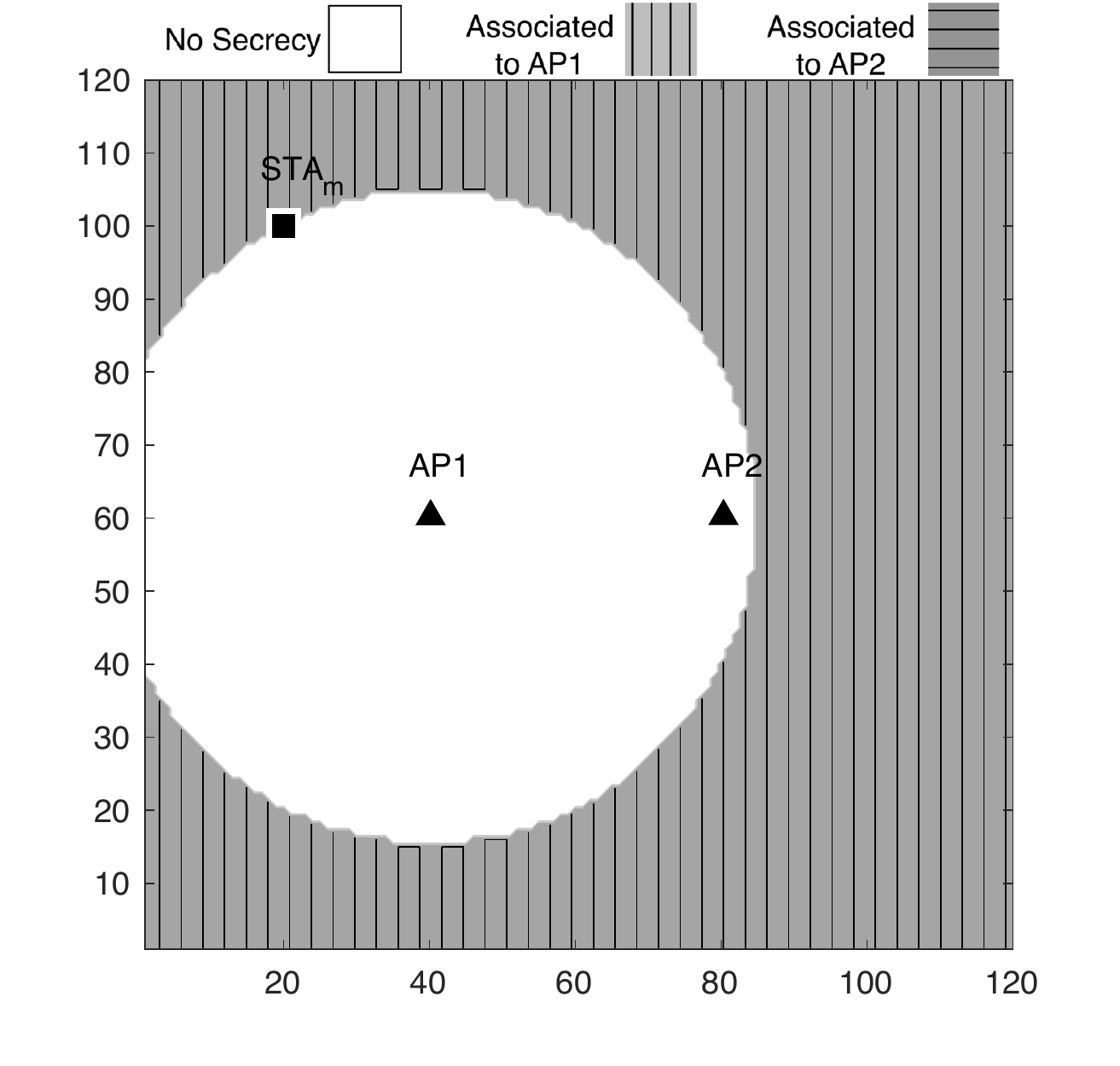}
         \vspace{-4mm}
         \caption{Association Map: Normal Wi-Fi}
         \label{fig:APmapWiFi1}
     \end{subfigure}
     \hfill
     \begin{subfigure}[b]{0.297\textwidth}
         \hspace{-7mm}
         \includegraphics[trim=0 0 0 0, clip,width=\textwidth]{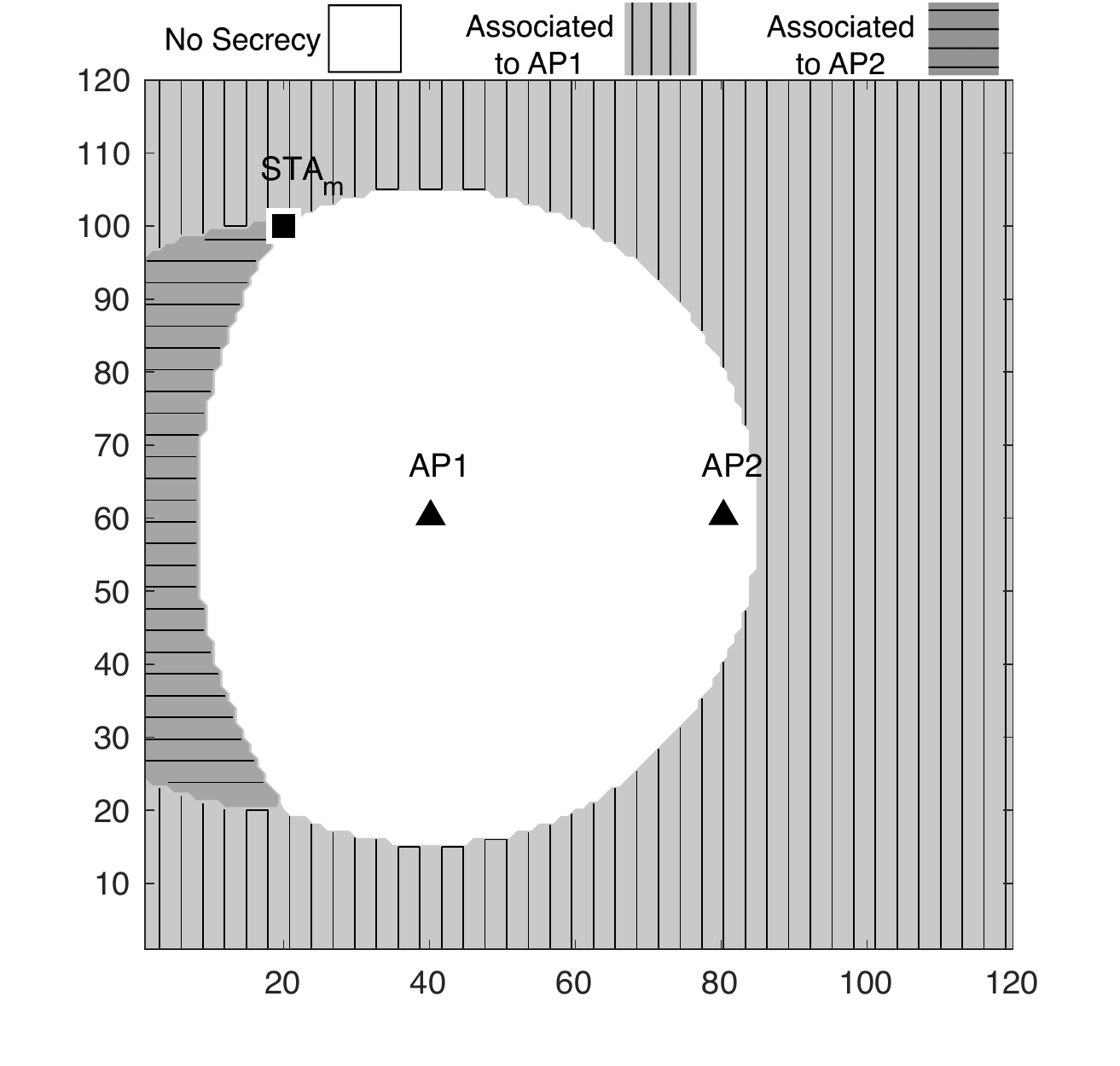}
         \vspace{-4mm}
         \caption{Association Map: Smart AP}
         \label{fig:APmapGC1}
     \end{subfigure}
     \hfill
     \begin{subfigure}[b]{0.297\textwidth}
         \hspace{-8.5mm}
         \includegraphics[trim=0 0 0 0, clip,width=\textwidth]{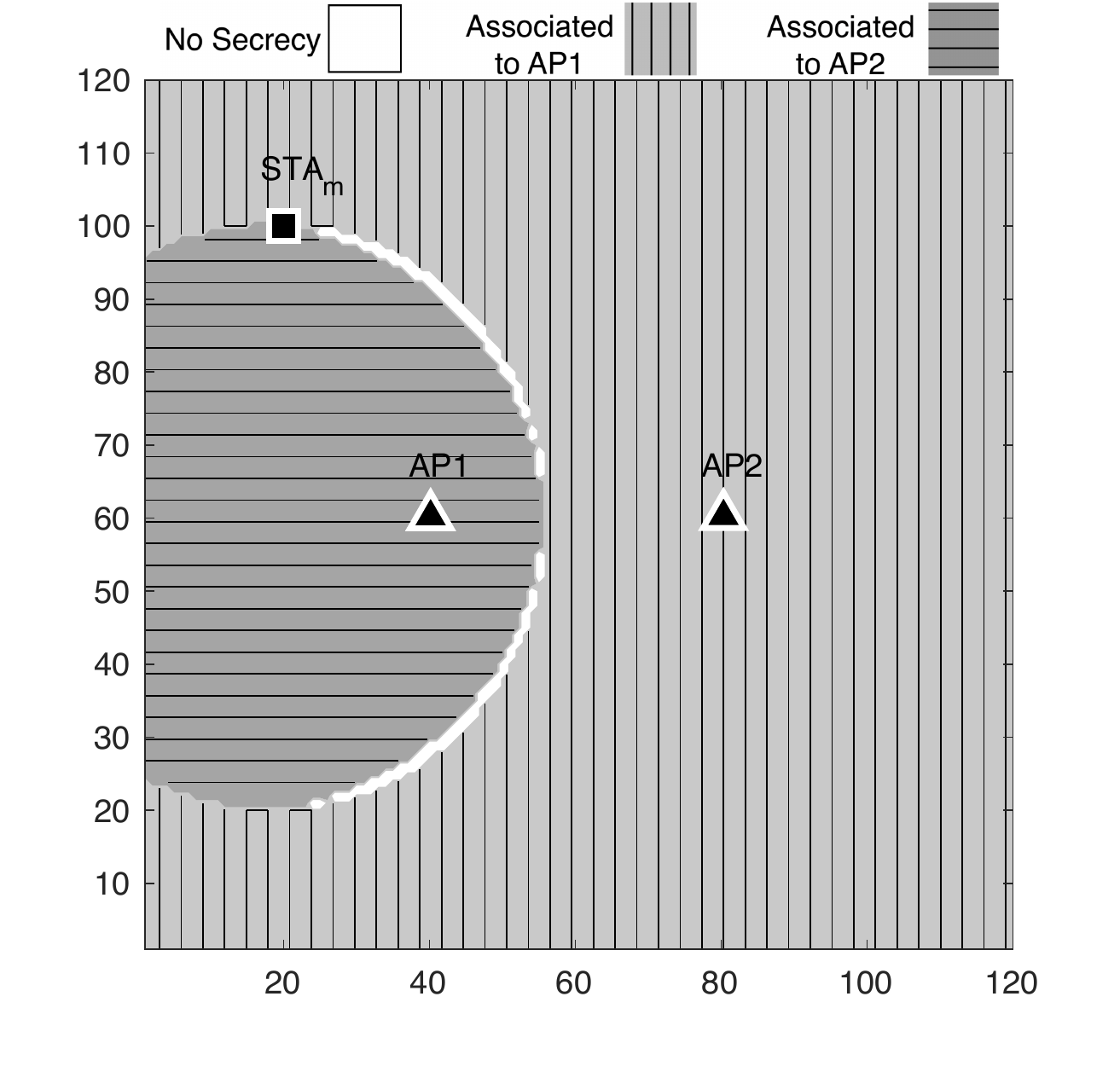}
         \vspace{-4mm}
         \caption{Association Map: Smart AP + FJ}
         \label{fig:APmapInterf1}
     \end{subfigure} \\

     \begin{subfigure}[b]{0.32\textwidth}
         \centering
         \includegraphics[trim=25 0 0 0, clip, width=\textwidth]{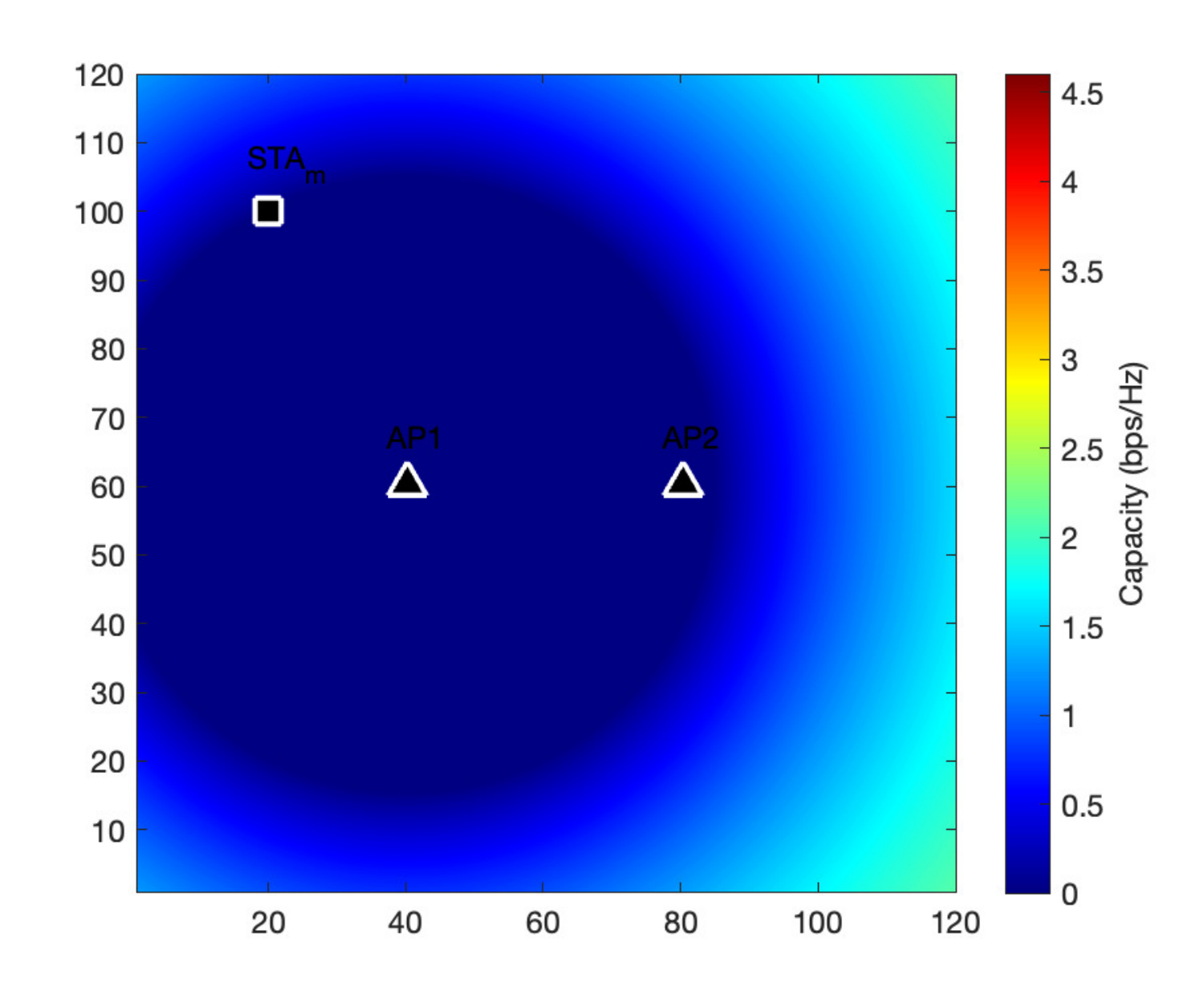}
         \vspace{-8mm}
         \caption{Secrecy Capacity: Normal Wi-Fi}
         \label{fig:CapSecWiFi1}
     \end{subfigure}
     \hfill
     \begin{subfigure}[b]{0.32\textwidth}
         \centering
         \includegraphics[trim=25 0 0 0, clip,width=\textwidth]{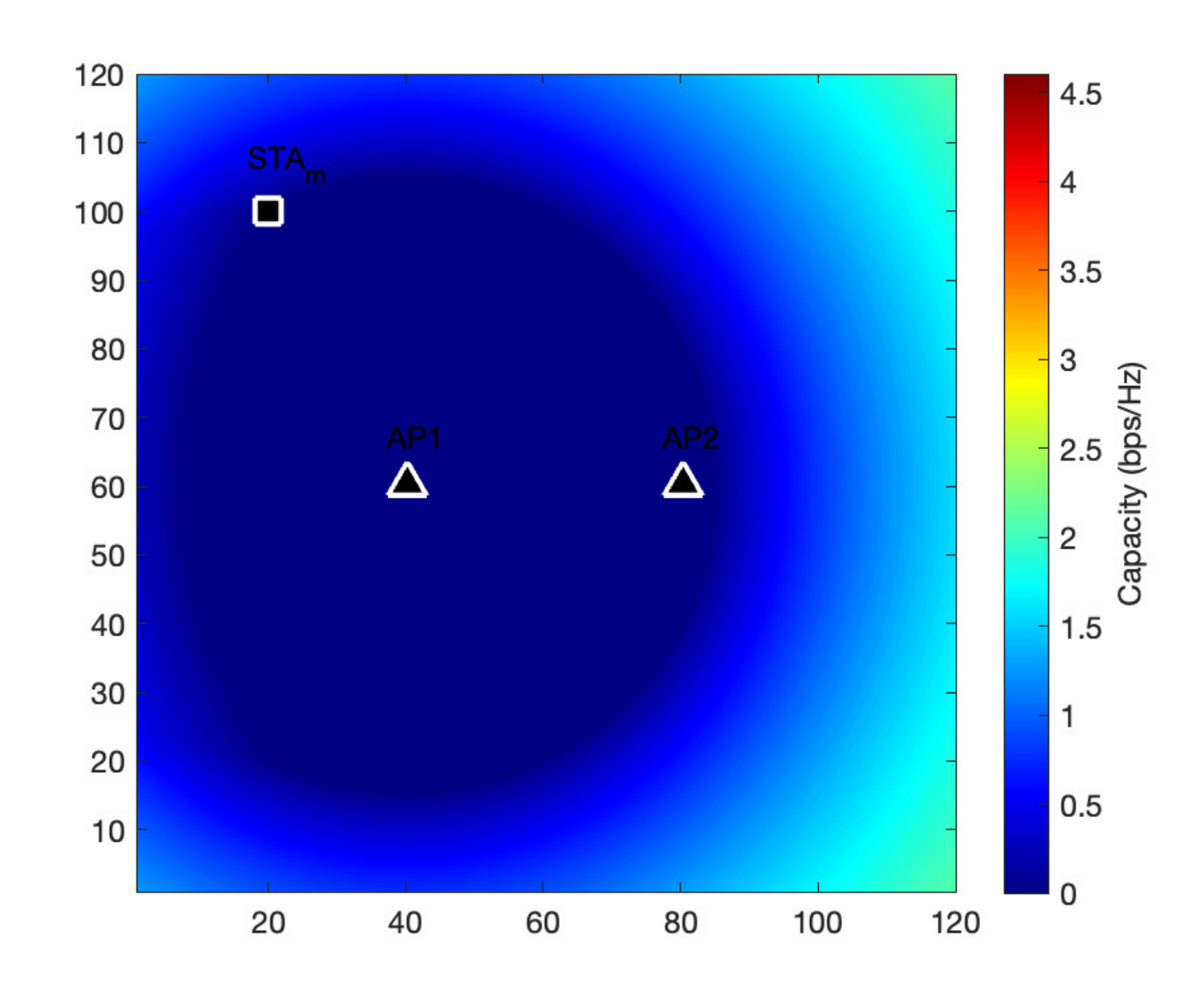}
         \vspace{-8mm}
         \caption{Secrecy Capacity: Smart AP}
         \label{fig:CapSecGC1}
     \end{subfigure}
     \hfill
     \begin{subfigure}[b]{0.32\textwidth}
         \centering
         \includegraphics[trim=25 0 0 0, clip,width=\textwidth]{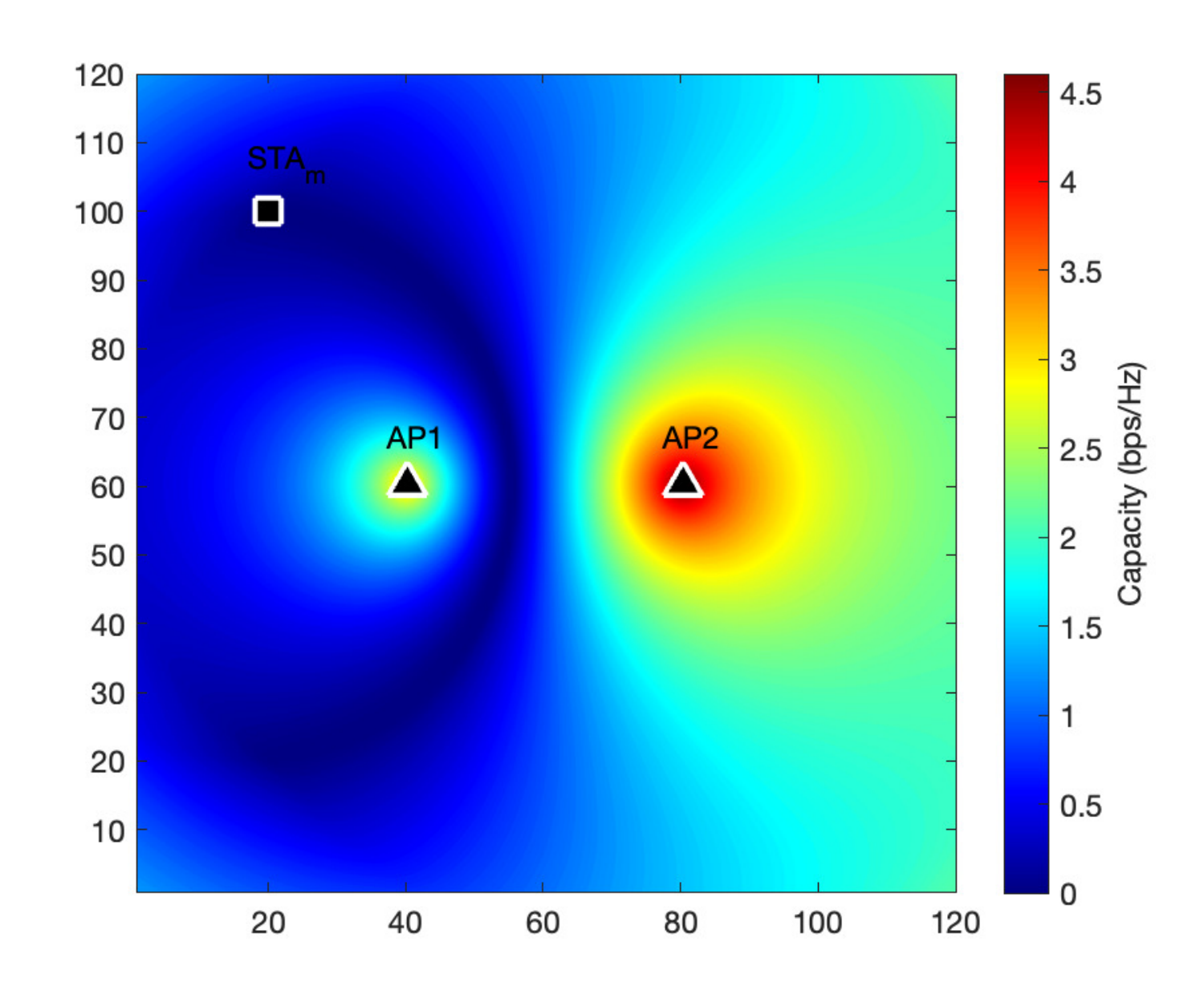}
         \vspace{-8mm}
         \caption{Secrecy Capacity: Smart AP + FJ}
         \label{fig:CapSecInterf1}
     \end{subfigure} 
     
        \caption{Association maps and STA$_m$ secrecy capacity for different locations of STA$_e$ where STA$_m$ is located at position $(20,100)$ (scenario 1), for three different Wi-Fi systems.}
        \label{fig:mapFigures1}
\end{figure*}

\begin{figure*}
    \centering
    \begin{subfigure}[]{0.297\textwidth}
        \hspace{-4.7mm}
        \includegraphics[trim=0 0 0 0, clip, width=\textwidth]{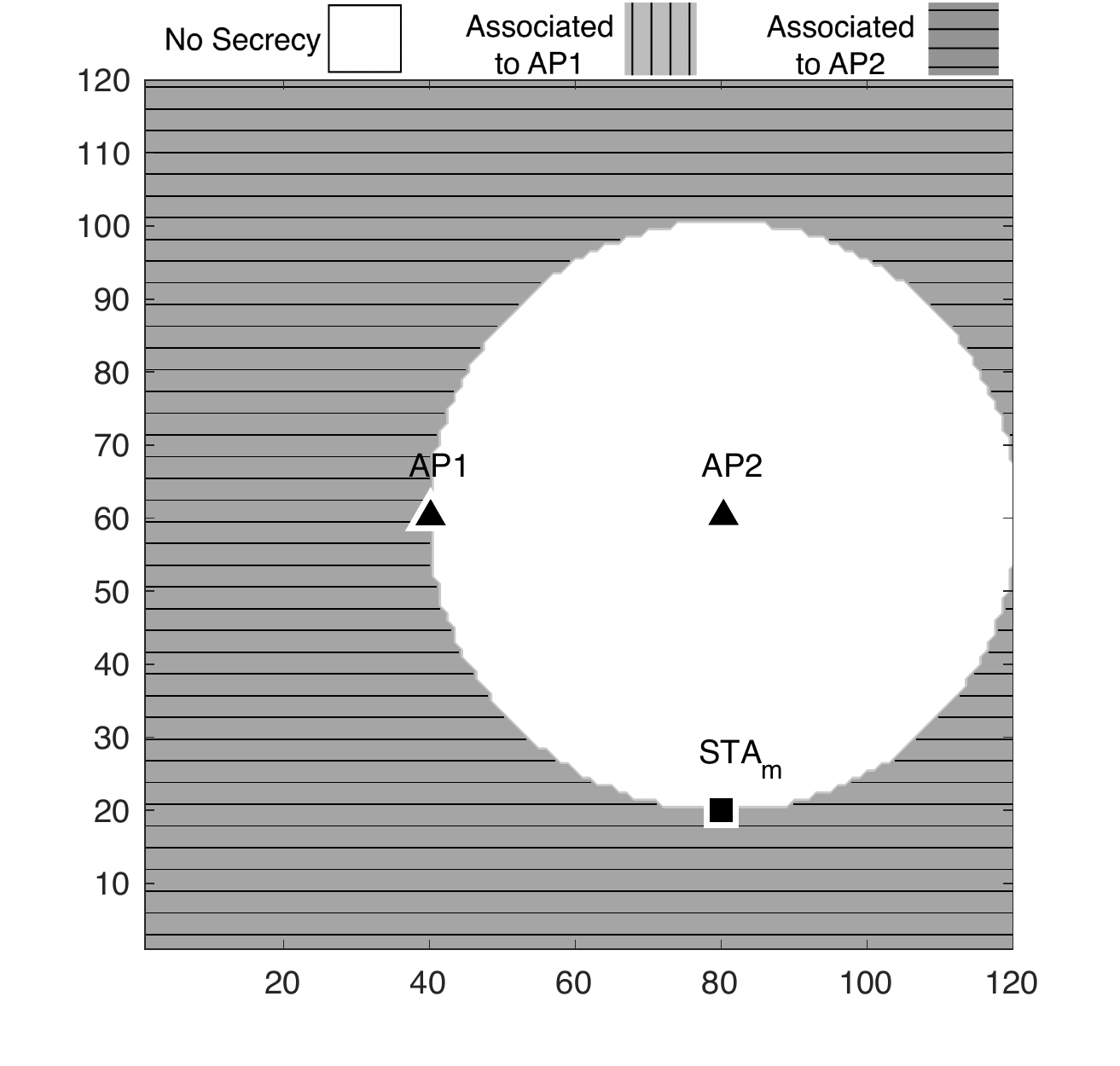}
        \vspace{-4mm}
        \caption{Association Map: Normal WiFi}
        \label{fig:APmapWiFi2}
    \end{subfigure}
    \hfill
    \begin{subfigure}[]{0.297\textwidth}
        \hspace{-7mm}
        \includegraphics[trim=0 0 0 0, clip,width=\textwidth]{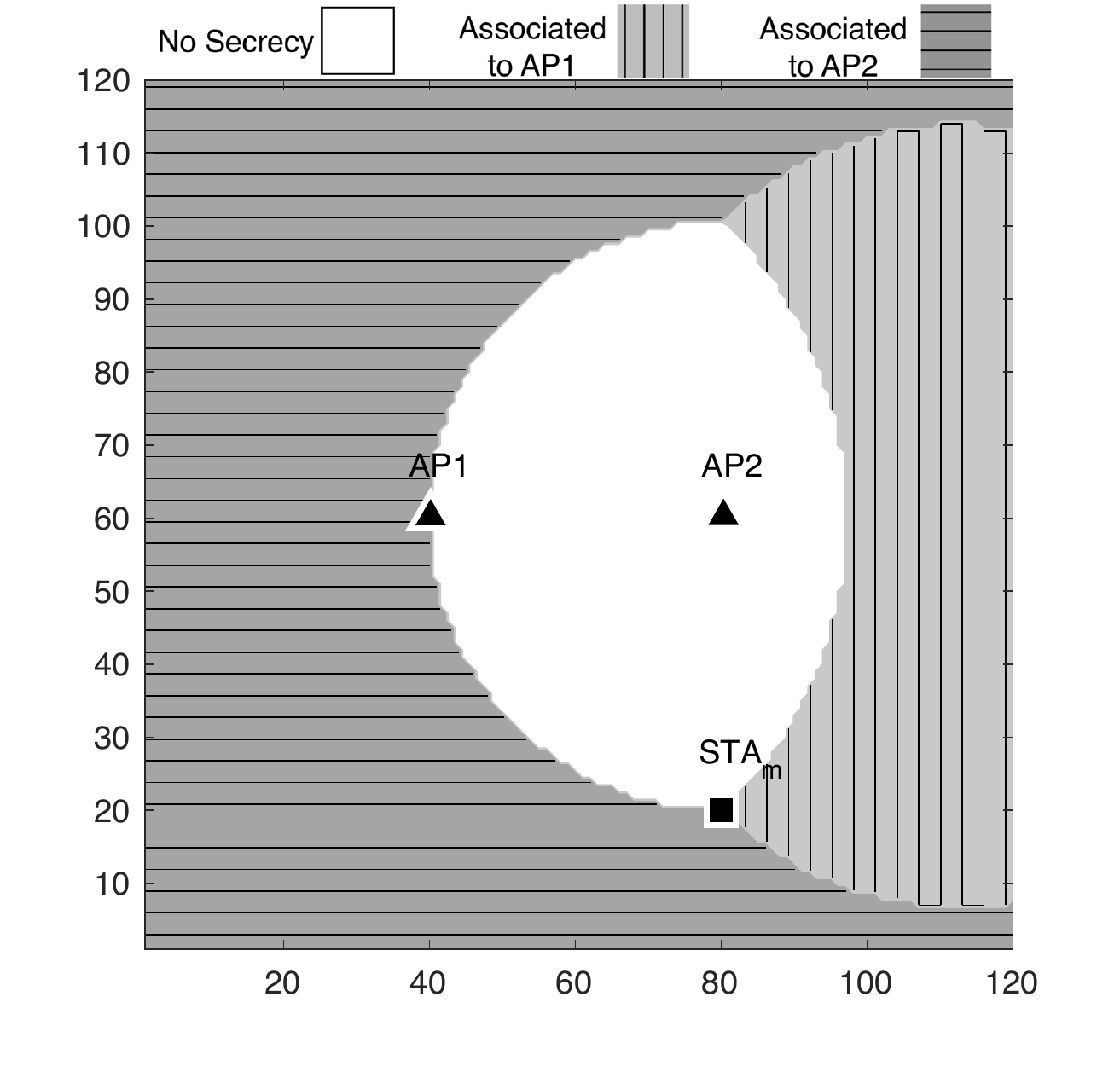}
        \vspace{-4mm}
        \caption{Association Map: Smart AP}
        \label{fig:APmapGC2}
    \end{subfigure}
    \hfill
    \begin{subfigure}[]{0.297\textwidth}
        \hspace{-8.5mm}
        \includegraphics[trim=0 0 0 0, clip,width=\textwidth]{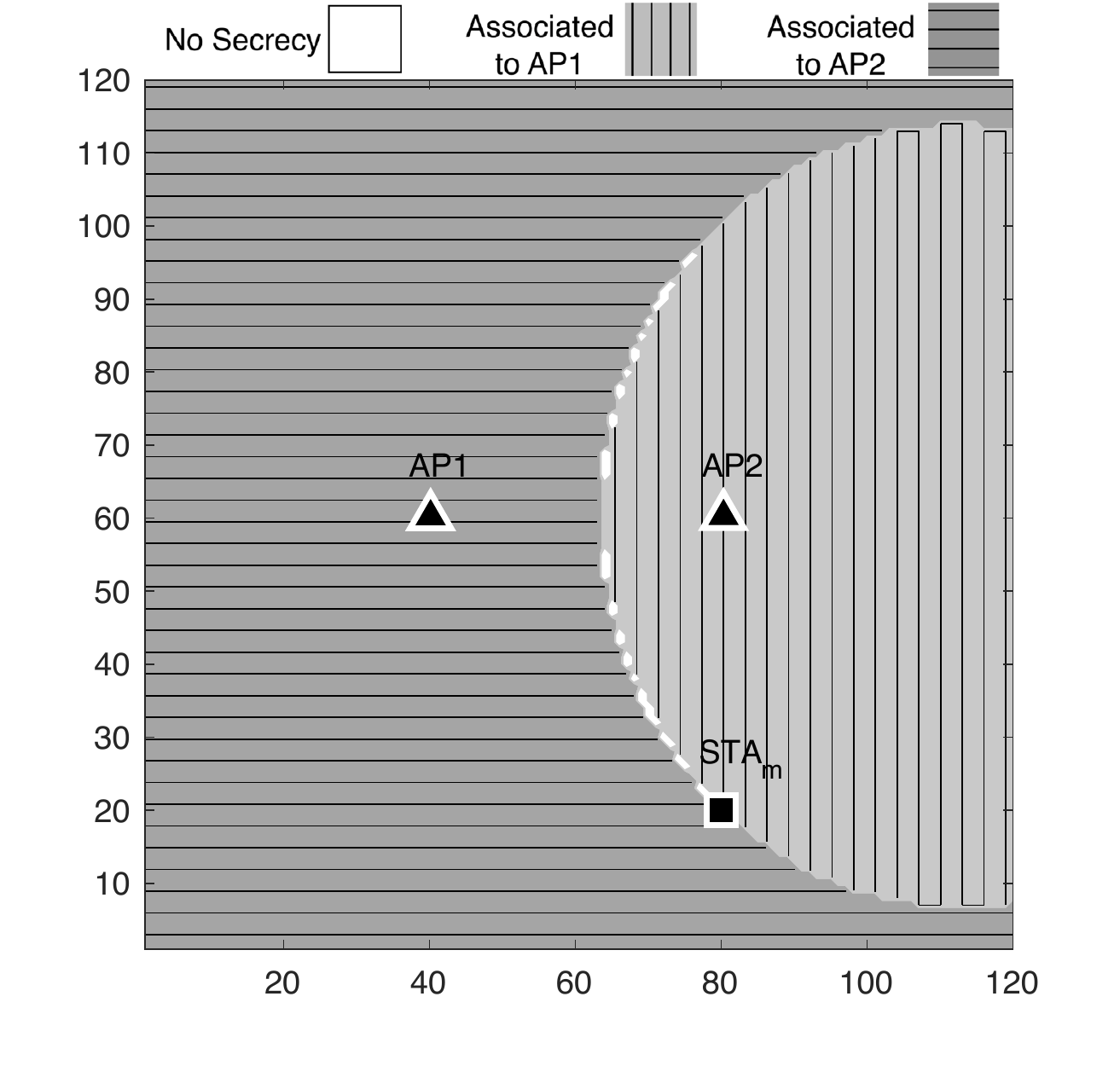}
        \vspace{-4mm}
        \caption{Association Map: Smart AP + FJ}
        \label{fig:APmapInterf2}
    \end{subfigure} \\

    \begin{subfigure}[]{0.32\textwidth}
        \centering
        \includegraphics[trim=25 0 0 0, clip, width=\textwidth]{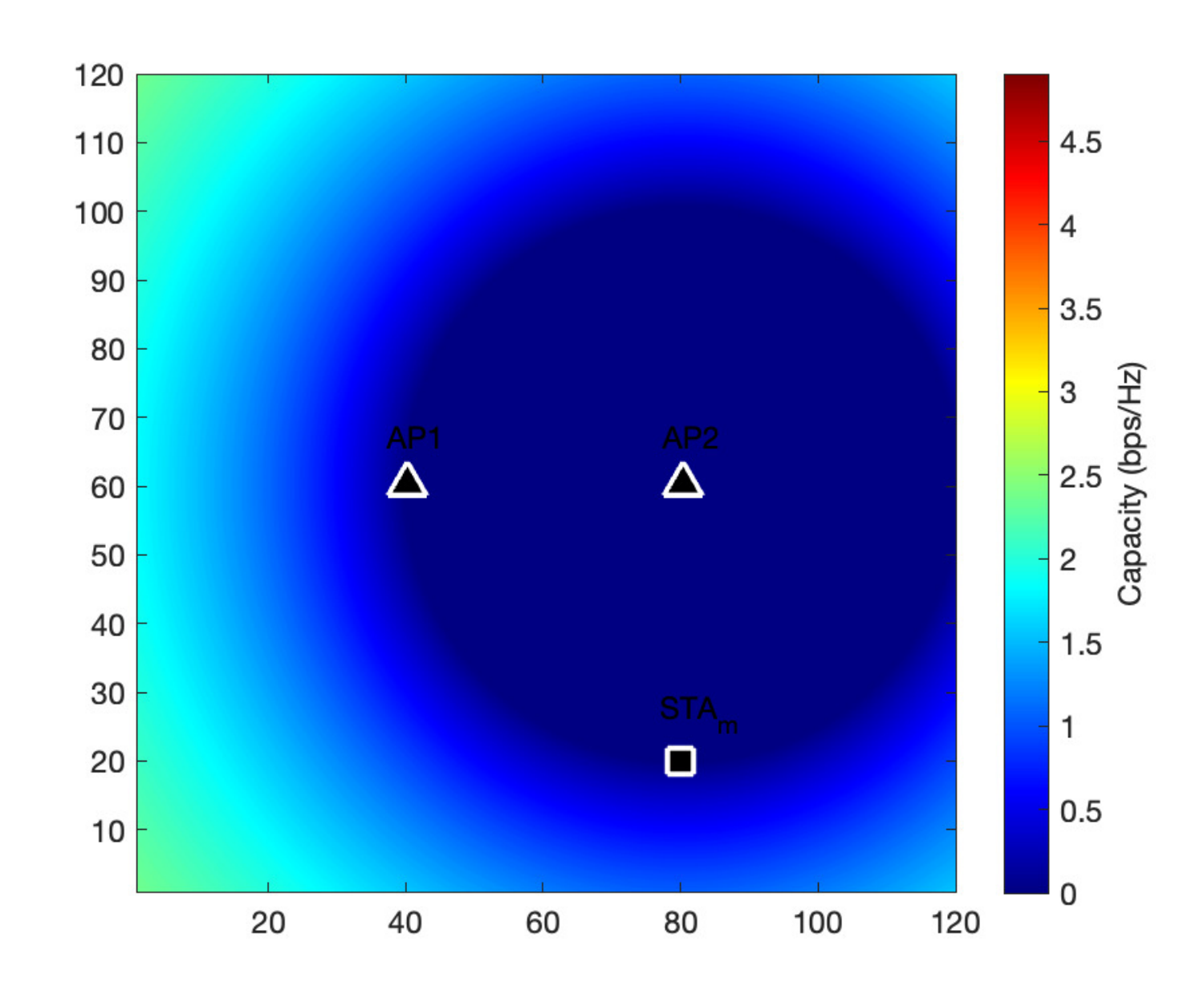}
        \vspace{-8mm}
        \caption{Secrecy Capacity: Normal Wi-Fi}
        \label{fig:CapSecWiFi2}
    \end{subfigure}
    \hfill
    \begin{subfigure}[]{0.32\textwidth}
        \centering
        \includegraphics[trim=25 0 0 0, clip,width=\textwidth]{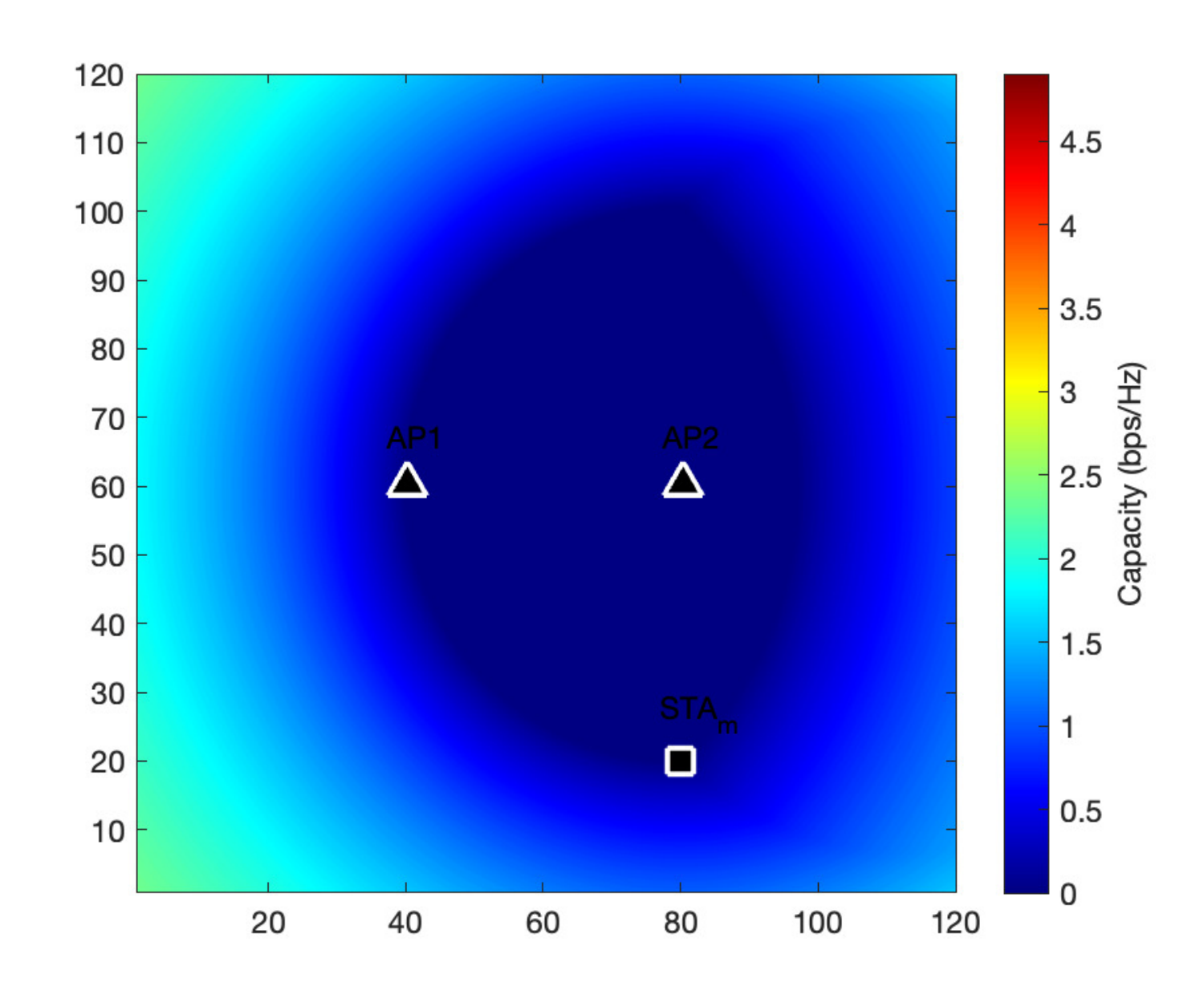}
        \vspace{-8mm}
        \caption{Secrecy Capacity: Smart AP}
        \label{fig:CapSecGC2}
    \end{subfigure}
    \hfill
    \begin{subfigure}[]{0.32\textwidth}
        \centering
        \includegraphics[trim=25 0 0 0, clip,width=\textwidth]{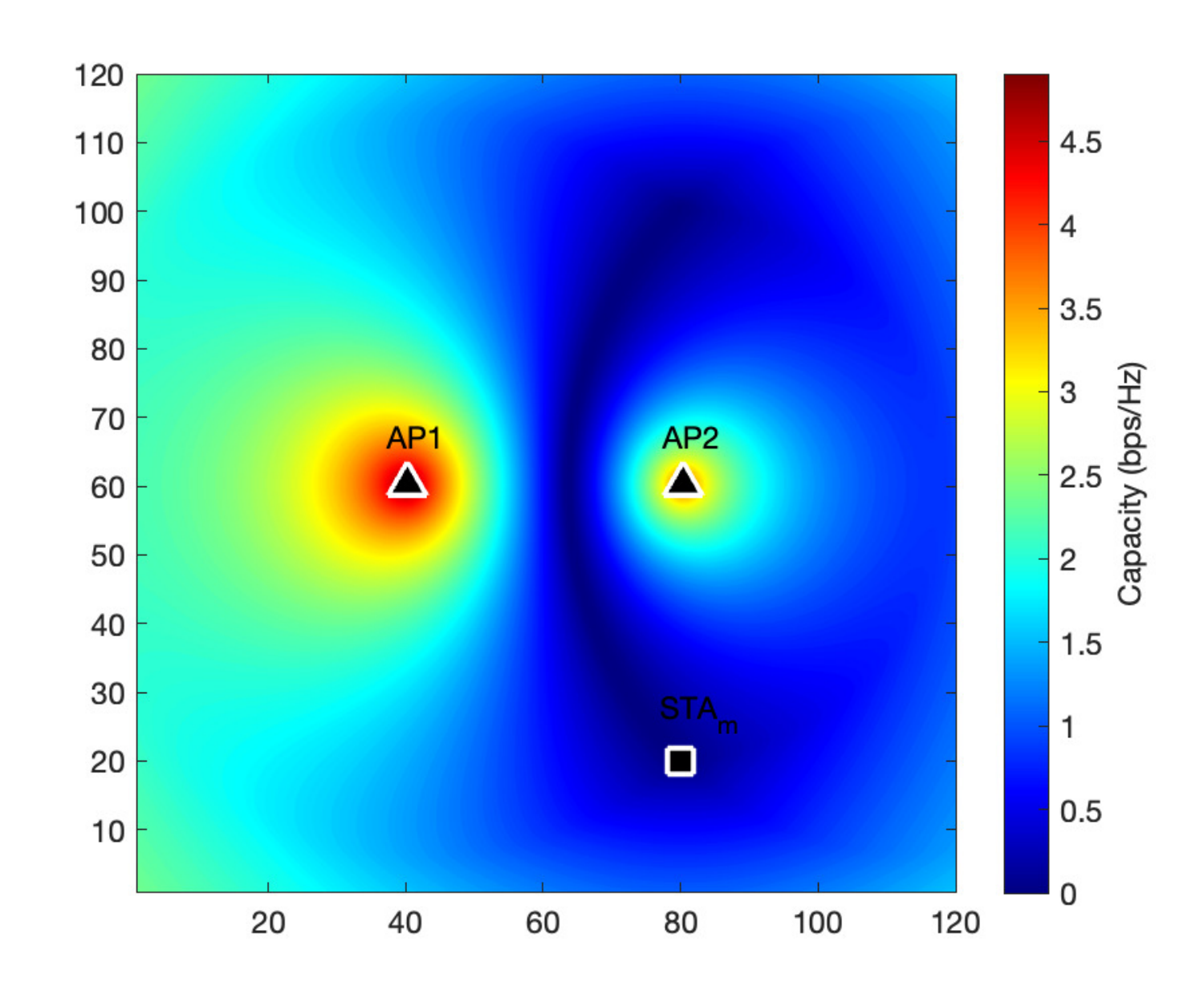}
        \vspace{-8mm}
        \caption{Secrecy Capacity: Smart AP + FJ}
        \label{fig:CapSecInterf2}
    \end{subfigure} 
    
        \caption{Scenario 2: STA$_m$ is located at position $(80, 20)$.}
        \label{fig:mapFigures2}
\end{figure*}

\begin{figure*}
    \centering
    \begin{subfigure}[b]{0.297\textwidth}
        \hspace{-4.7mm}
        \includegraphics[trim=0 0 0 0, clip, width=\textwidth]{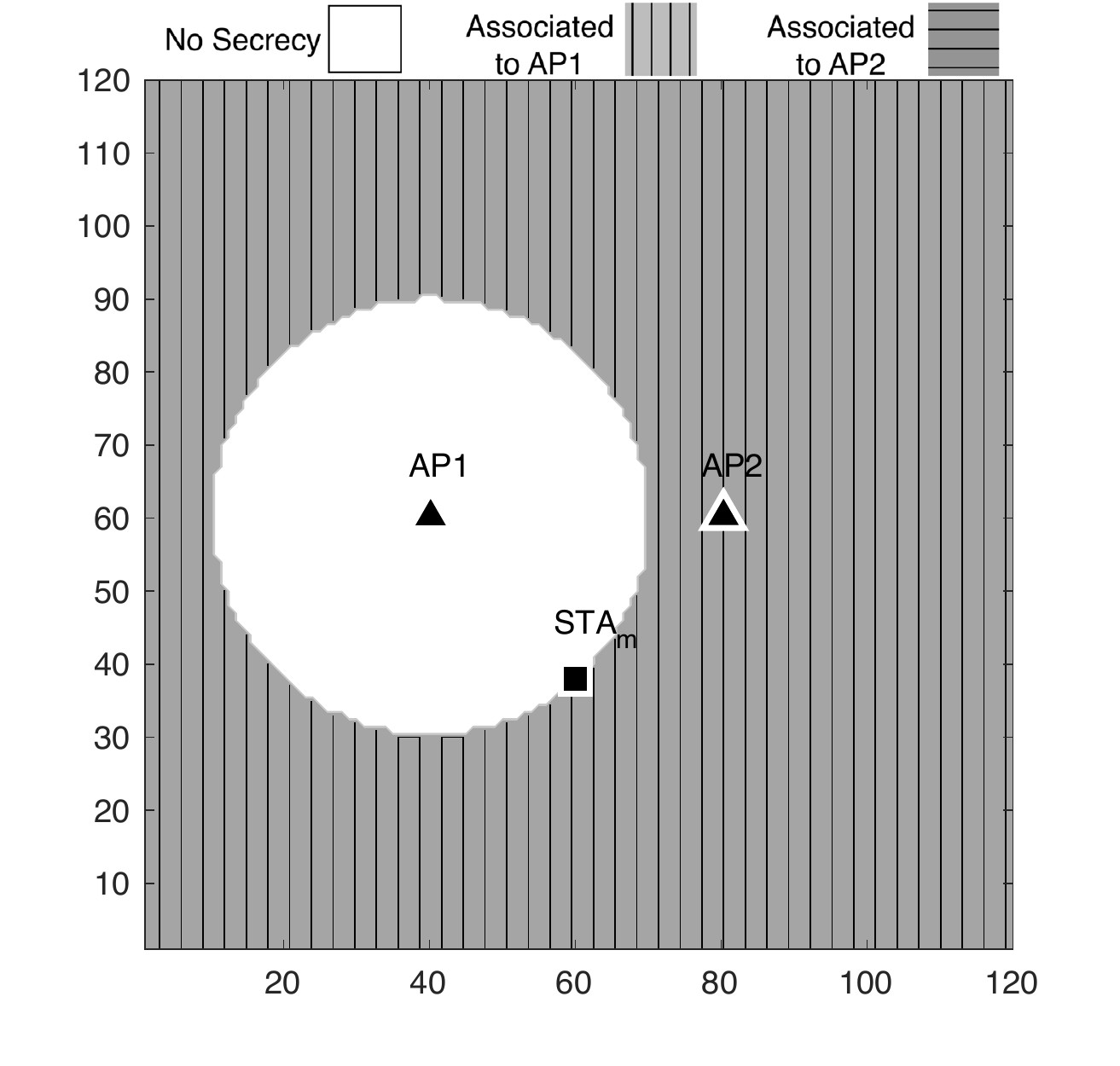}
        \vspace{-4mm}
        \caption{Association Map: Normal Wi-Fi}
        \label{fig:APmapWiFi3}
    \end{subfigure}
    \hfill
    \begin{subfigure}[b]{0.297\textwidth}
        \hspace{-7mm}
        \includegraphics[trim=0 0 0 0, clip,width=\textwidth]{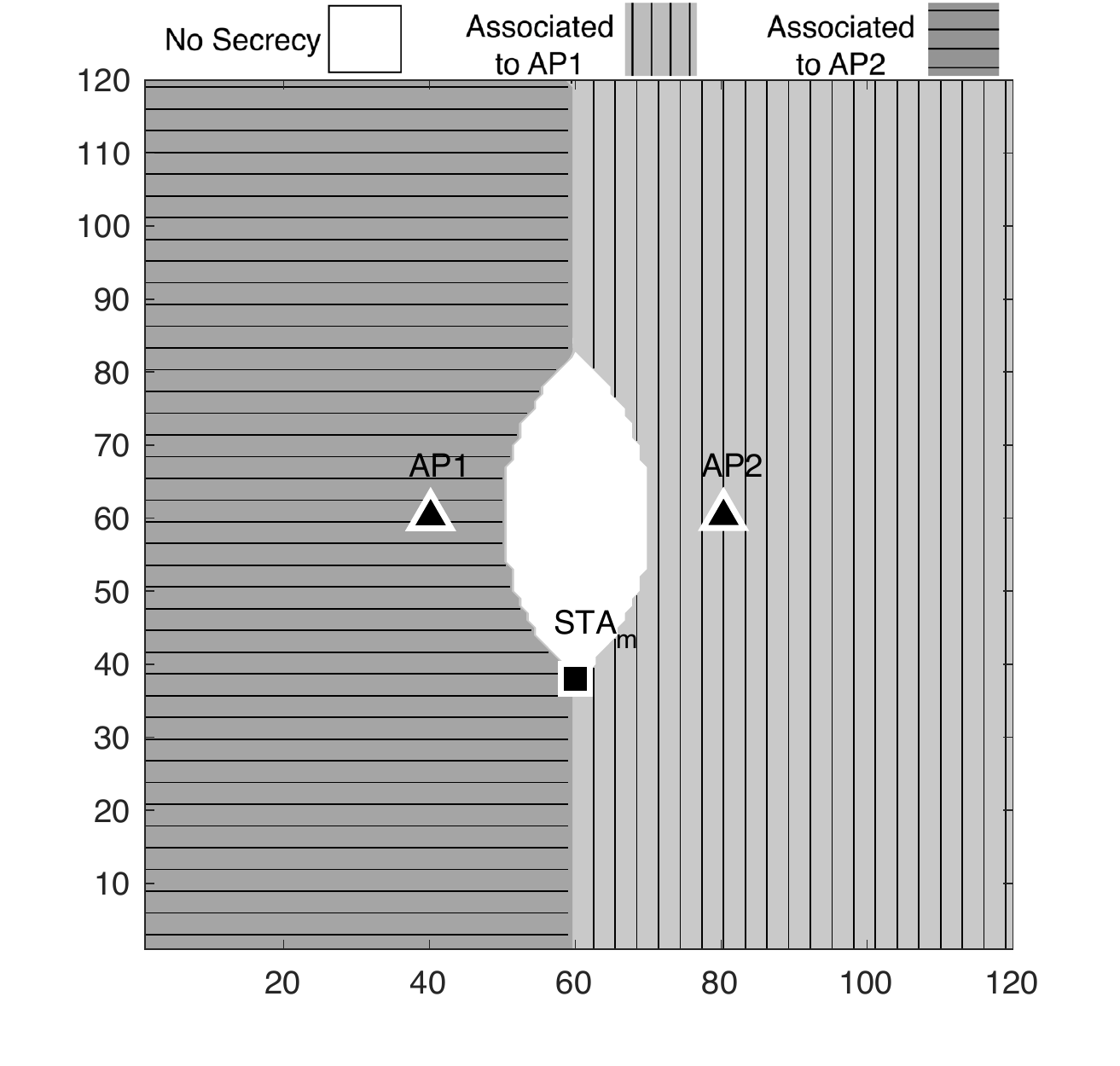}
        \vspace{-4mm}
        \caption{Association Map: Smart AP}
        \label{fig:APmapGC3}
    \end{subfigure}
    \hfill
    \begin{subfigure}[b]{0.297\textwidth}
        \hspace{-8.5mm}
        \includegraphics[trim=0 0 0 0, clip,width=\textwidth]{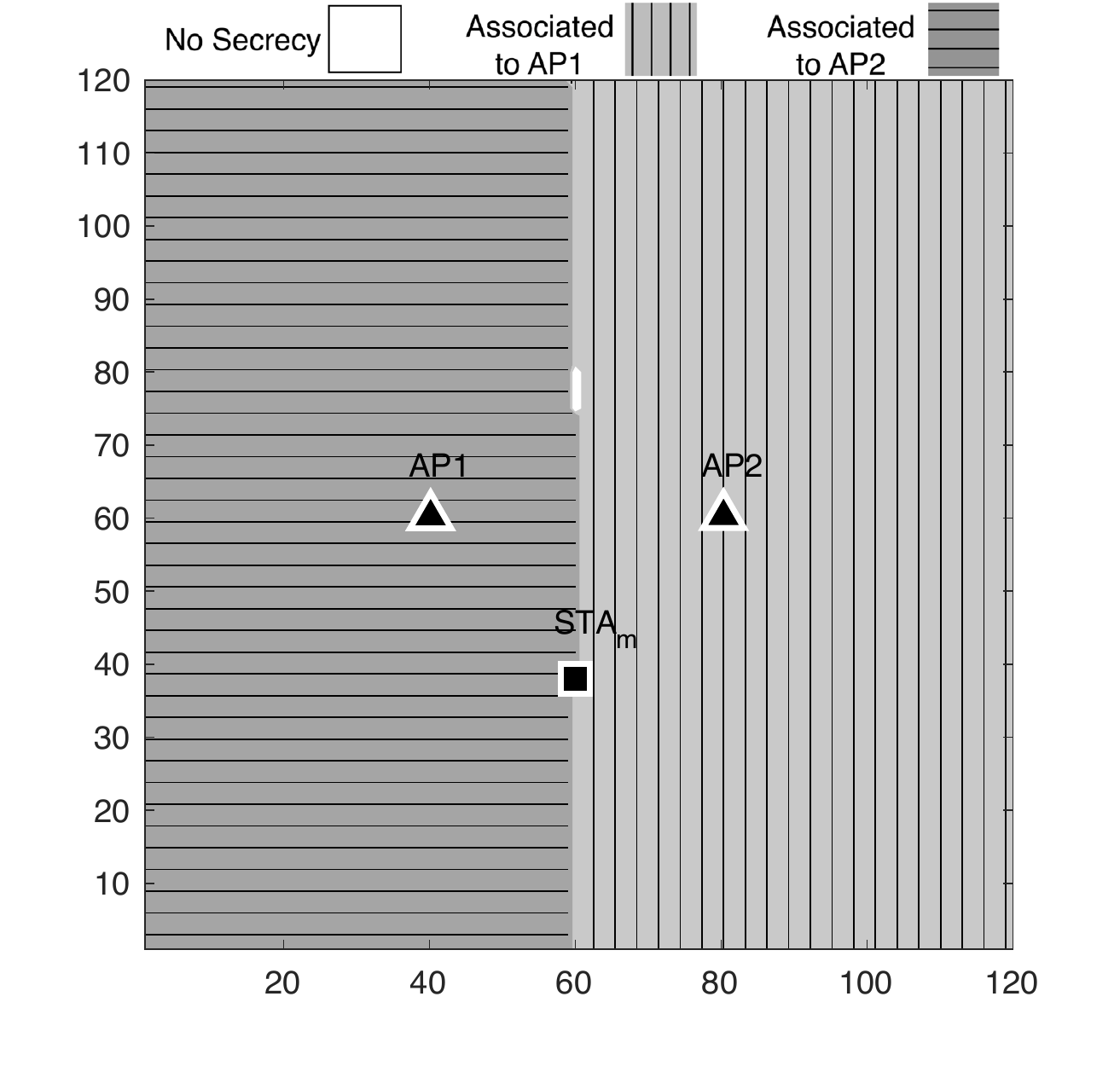}
        \vspace{-4mm}
        \caption{Association Map: Smart AP + FJ}
        \label{fig:APmapInterf3}
    \end{subfigure} \\

    \begin{subfigure}[b]{0.32\textwidth}
        \centering
        \includegraphics[trim=25 0 0 0, clip, width=\textwidth]{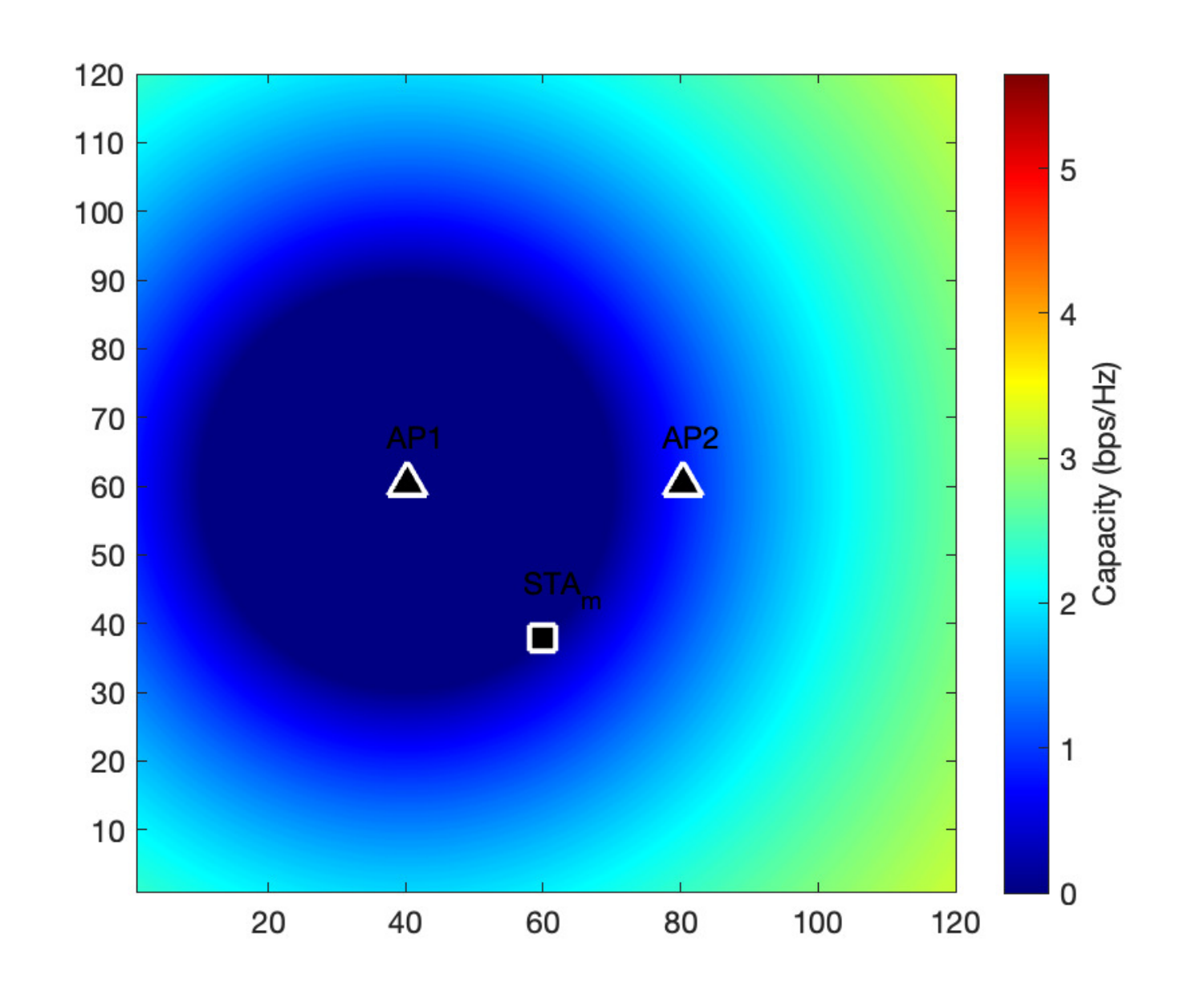}
        \vspace{-8mm}
        \caption{Secrecy Capacity: Normal Wi-Fi}
        \label{fig:CapSecWiFi3}
    \end{subfigure}
    \hfill
    \begin{subfigure}[b]{0.32\textwidth}
        \centering
        \includegraphics[trim=25 0 0 0, clip,width=\textwidth]{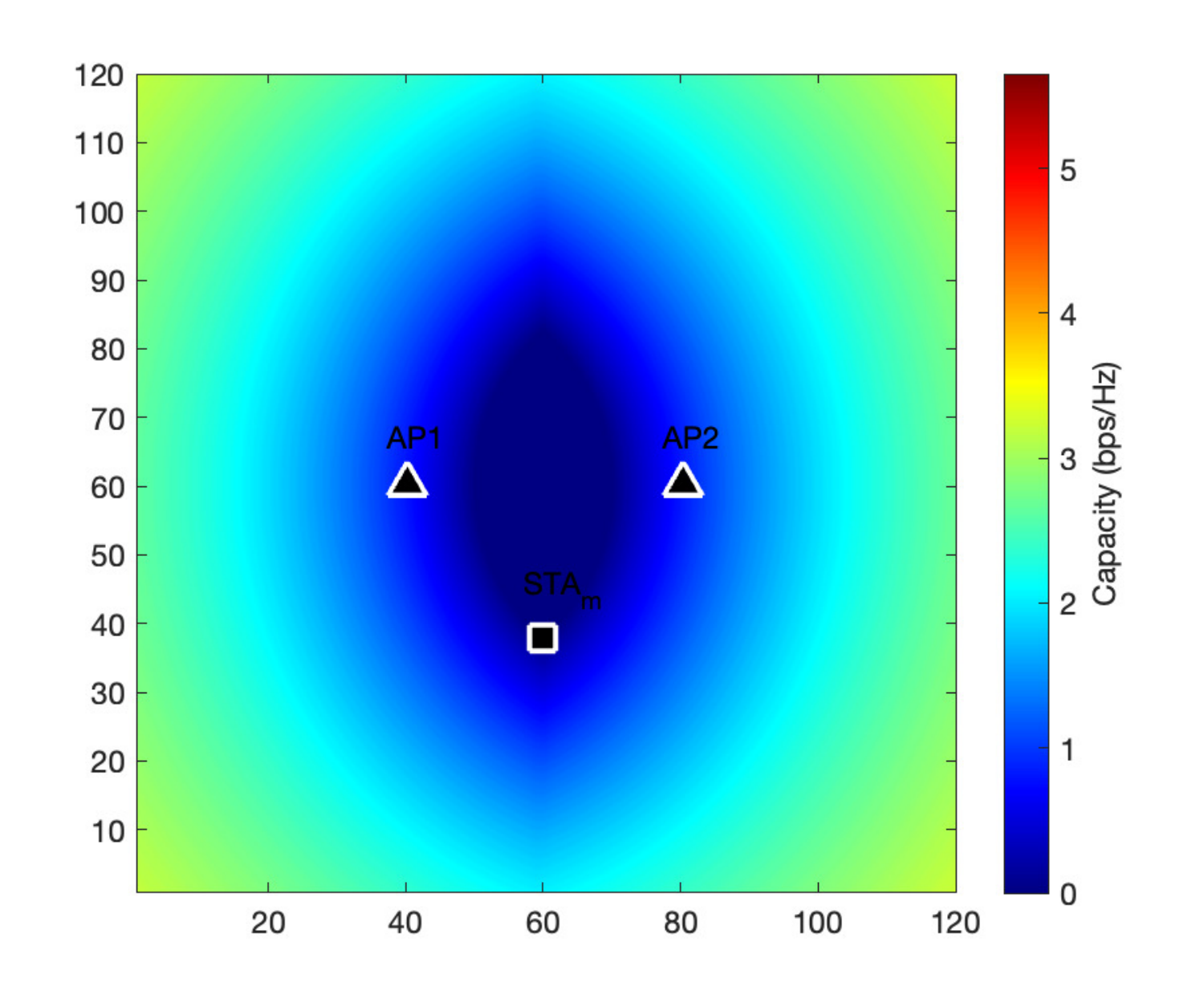}
        \vspace{-8mm}
        \caption{Secrecy Capacity: Smart AP}
        \label{fig:CapSecGC3}
    \end{subfigure}
    \hfill
    \begin{subfigure}[b]{0.32\textwidth}
        \centering
        \includegraphics[trim=25 0 0 0, clip,width=\textwidth]{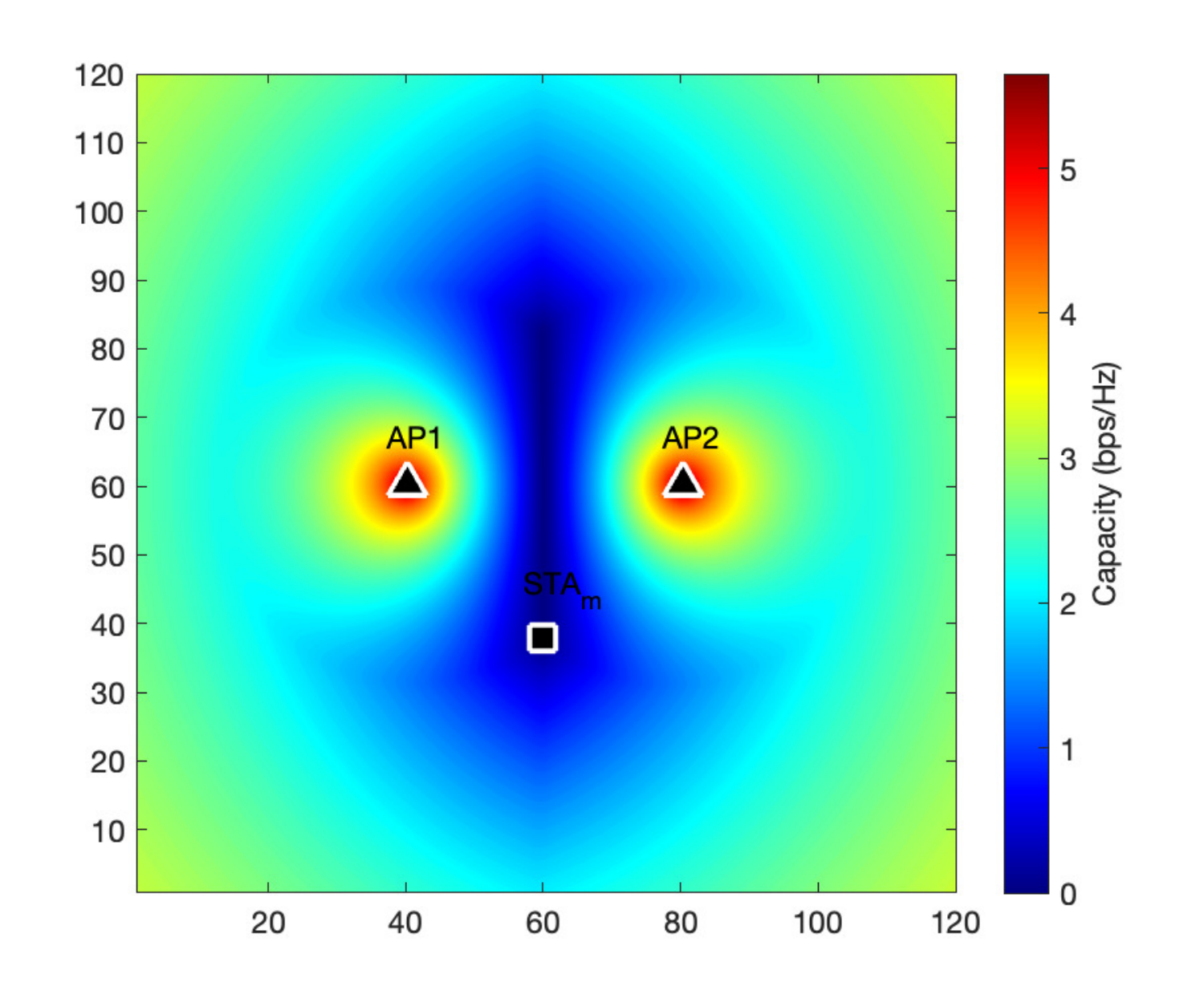}
        \vspace{-8mm}
        \caption{Secrecy Capacity: Smart AP + FJ}
        \label{fig:CapSecInterf3}
    \end{subfigure} 
    
        \caption{Scenario 3: STA$_m$ is located at position $(60, 38)$. }
        \label{fig:mapFigures3}
\end{figure*}

We consider a $120m \times 120m$ environment and all coordinates are expressed in meter (m). The two APs are located at positions $(40,60)$ and $(80,60)$ and operate at $f_0 = 2.4$ GHz. The  AP associated to STA$_m$ uses a fixed transmit power of $P_i = 50$ mW. When the idle AP$_j$ is used to introduce FJ, we assume $P_{t_n}^{\max} = 50$ mW. The noise power at both STA$_m$ and STA$_e$ is  $N = N_m = N_e = -70$ dBm $= 10^{-10}$ W. A path loss exponent of $\alpha = 2$ and a reference distance $d_0 = 1m$ are assumed for the entire map. 
Then, for a fixed location of STA$_m$, we run the simulation and for each potential location of STA$_e$, we first select the AP to associate with STA$_m$ based on one of the above three methods and then calculate the secrecy capacity $C_{i,m}-C_{i,e}$, eavesdropping capacity $C_{i,e}$, and the coverage ratio. The latter is the ratio of the area with positive secrecy capacity and the total area of the map, i.e. all locations where the eavesdropper may be located.  We repeat the simulations for three different scenarios where STA$_m$ is located at position $(20,100)$ (scenario 1), $(80, 20)$ (scenario 2) and $(60, 38)$ (scenario 3). 
The results are visualized as a map in \figurename~\ref{fig:mapFigures1}, \ref{fig:mapFigures2} and \ref{fig:mapFigures3}, respectively.

 For scenario 1, \figurename~\ref{fig:APmapWiFi1} shows for the ``normal Wi-Fi" how STA$_m$ associates to the AP with the highest received SINR, which is the nearest one. If  STA$_e$ positions itself closer to this AP$_i$, it will be located in the white area, which means that no secrecy can be achieved for STA$_m$. In \figurename~\ref{fig:CapSecWiFi1} this shows as a secrecy capacity of $0$ (zero), i.e. dark blue. Outside this circle, the secrecy capacity is greater than zero. \figurename~\ref{fig:APmapGC1} and \ref{fig:CapSecGC1} illustrate the results for the AP selection algorithm of \cite{Faycal2020globecom}. The type of shading (horizontal or vertical lines) indicates which AP STA$_m$ is connected to, given the location of STA$_e$, such that the secrecy capacity is maximized. Again, if STA$_e$ is in the white area, no secrecy can be achieved, regardless of which AP STA$_m$ is connected to. But this area is now significantly smaller than in \figurename~\ref{fig:APmapWiFi1}, showing the effectiveness of the algorithm in \cite{Faycal2020globecom}. When we now add FJ, the white area almost disappears, as shown in  \figurename~\ref{fig:APmapInterf1}. This means that, for this scenario, secrecy can be achieved almost everywhere. \figurename~\ref{fig:CapSecInterf1} shows that, in this case, the secrecy capacity for STA$_m$ has also increased significantly for most locations of STA$_e$. 

Results for two other locations of STA$_m$ (i.e. scenario 2 and scenario 3) are presented in \figurename~\ref{fig:mapFigures2} and \ref{fig:mapFigures3}. From these results we can draw the same conclusions as for scenario 1.

\begin{figure*}[htb]
     \centering
     \begin{subfigure}[b]{0.32\textwidth}
         \centering
         \includegraphics[trim=25 0 0 0, clip, width=\textwidth]{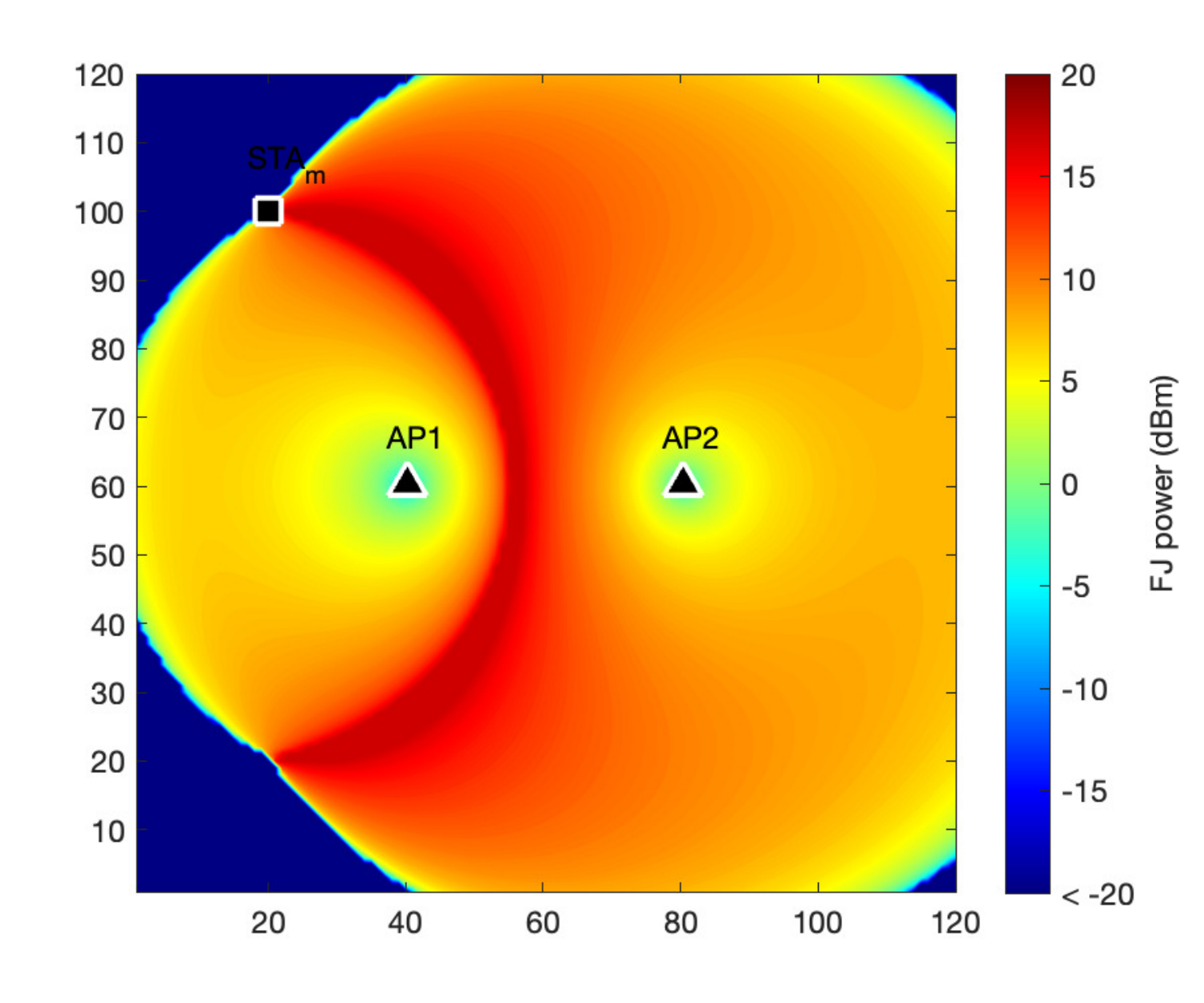}
         \vspace{-8mm}
         \caption{Scenario 1}
         \label{fig:InterferenceMap1}
     \end{subfigure}
     \hfill
     \begin{subfigure}[b]{0.32\textwidth}
         \centering
         \includegraphics[trim=25 0 0 0, clip,width=\textwidth]{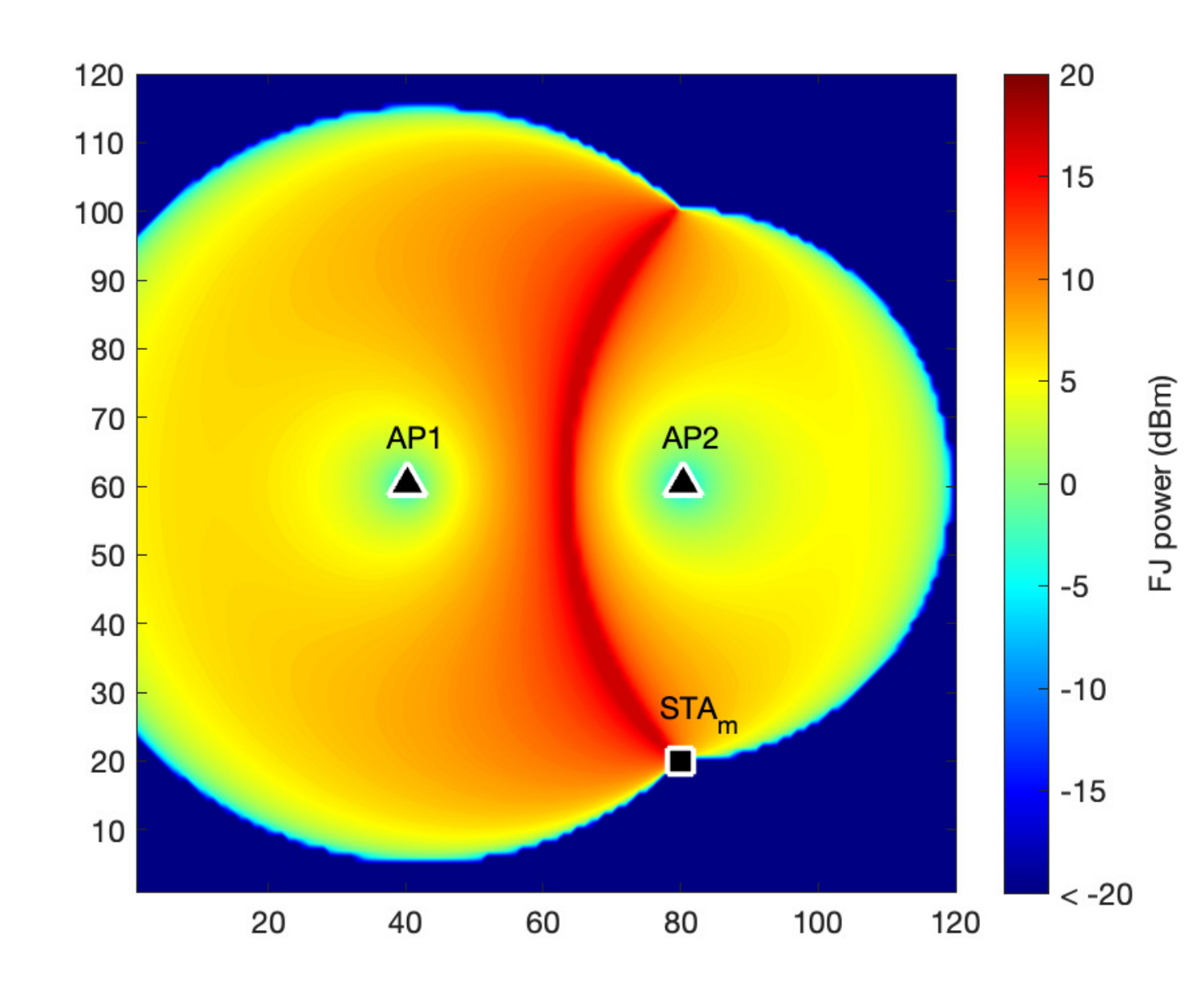}
         \vspace{-8mm}
         \caption{Scenario 2}
         \label{fig:InterferenceMap2}
     \end{subfigure}
     \hfill
     \begin{subfigure}[b]{0.32\textwidth}
         \centering
         \includegraphics[trim=25 0 0 0, clip,clip,width=\textwidth]{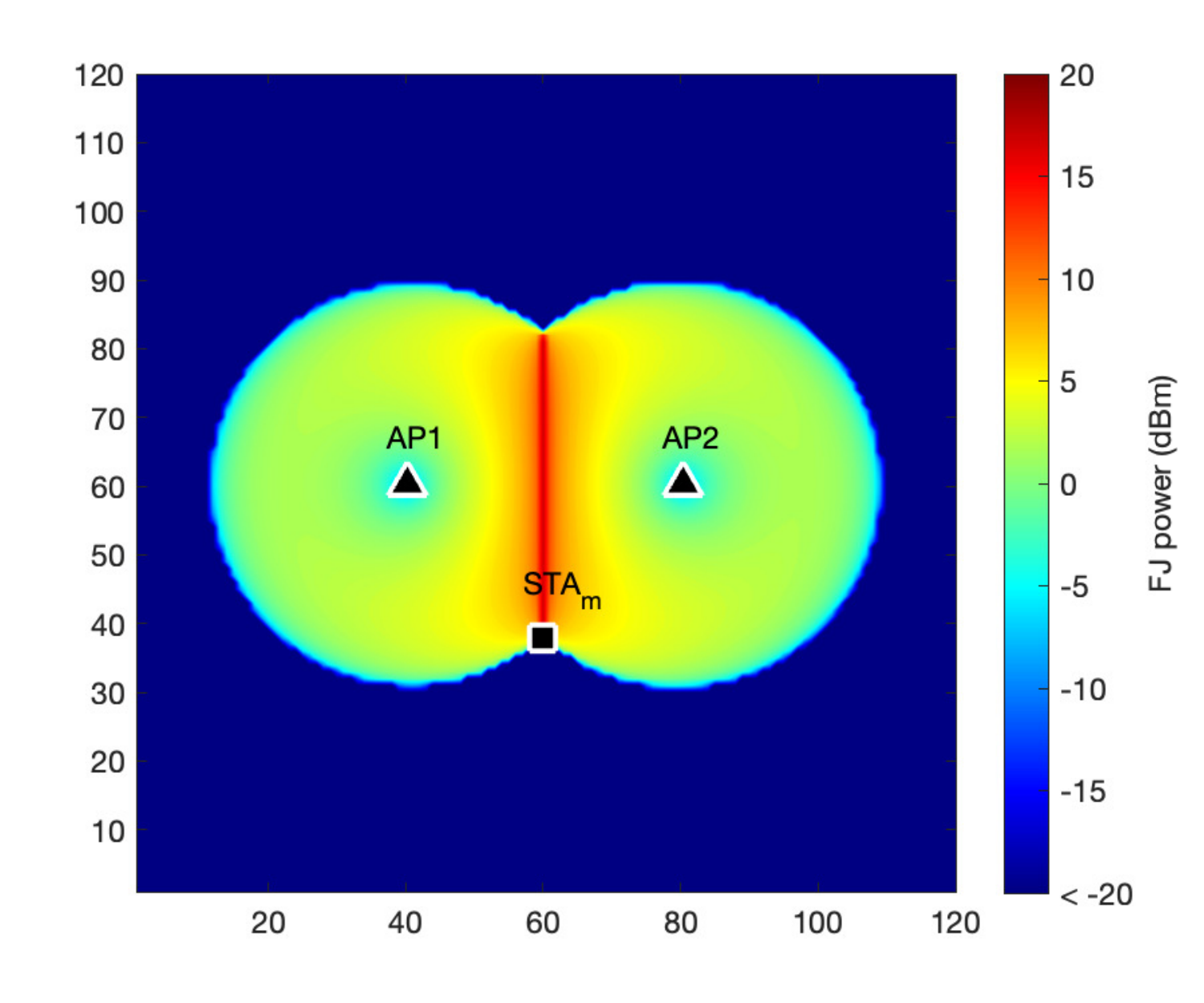}
         \vspace{-8mm}
         \caption{Scenario 3}
         \label{fig:InterferenceMap3}
     \end{subfigure} 
        \caption{FJ power generated by the idle AP for any given eavesdropper location.}
        \label{fig:InterferenceMap}
\end{figure*}

For all three scenarios, \figurename~\ref{fig:InterferenceMap} illustrates the FJ power that is generated by AP$_j$ to optimize the capacity and enhance the secrecy coverage. In our simulations, given a constant transmit power of 17 dBm for AP$_i$, the FJ power of AP$_j$ 
is about 5-10 dBm for the most of potential locations of the eavesdropper to optimize the secrecy capacity for STA$_m$. In contrast, optimal secrecy can be achieved without jamming when the eavesdropper is located on the edge area of the map.

Next, for every scenario and for every algorithm, we calculated the average secrecy capacity and the average eavesdropper's capacity for all possible locations of STA$_e$. Let STA$_m$ and STA$_e$ be located at ($x_m, y_m$) and ($x, y$), respectively. The channel capacity of STA$_m$ and STA$_e$ can then be represented as $C_{i,m}(x,y|x_m,y_m)$ and $C_{i,e}(x,y|x_m,y_m)$, respectively, where $x,y \in\{1,2, ... , K\}$, i.e. we vary the location of STA$_e$ in steps of 1 m in horizontal or vertical directions. This yields the average secrecy capacity

\begin{align*}
   E({C_{sec,m}})&= 
   \frac{1}{K^2}\sum_{x=1}^{K}\sum_{y=1}^{K} \big(C_{i,m}(x,y|x_m,y_m)\\ &\qquad \qquad \qquad- C_{i,e}(x,y|x_m,y_m)\big),
   \numberthis \label{eq:avg_sec_cap}
\end{align*}

with $K=120$. In the same way, we calculate the average eavesdropping capacity as
\begin{equation}
    E({C_{e}})=\frac{1}{K^2}\sum_{x=1}^{K}\sum_{y=1}^{K} C_{i,e}(x,y|x_m,y_m).
    \label{eq:avg_eav_cap}
\end{equation}

We then repeated the simulation for 10,000 random locations of STA$_m$. 

\begin{figure}[htb]
     \centering
     \begin{subfigure}[b]{0.5\textwidth}
         \centering
        \includegraphics[trim=55 0 0 0, clip,width=\textwidth]{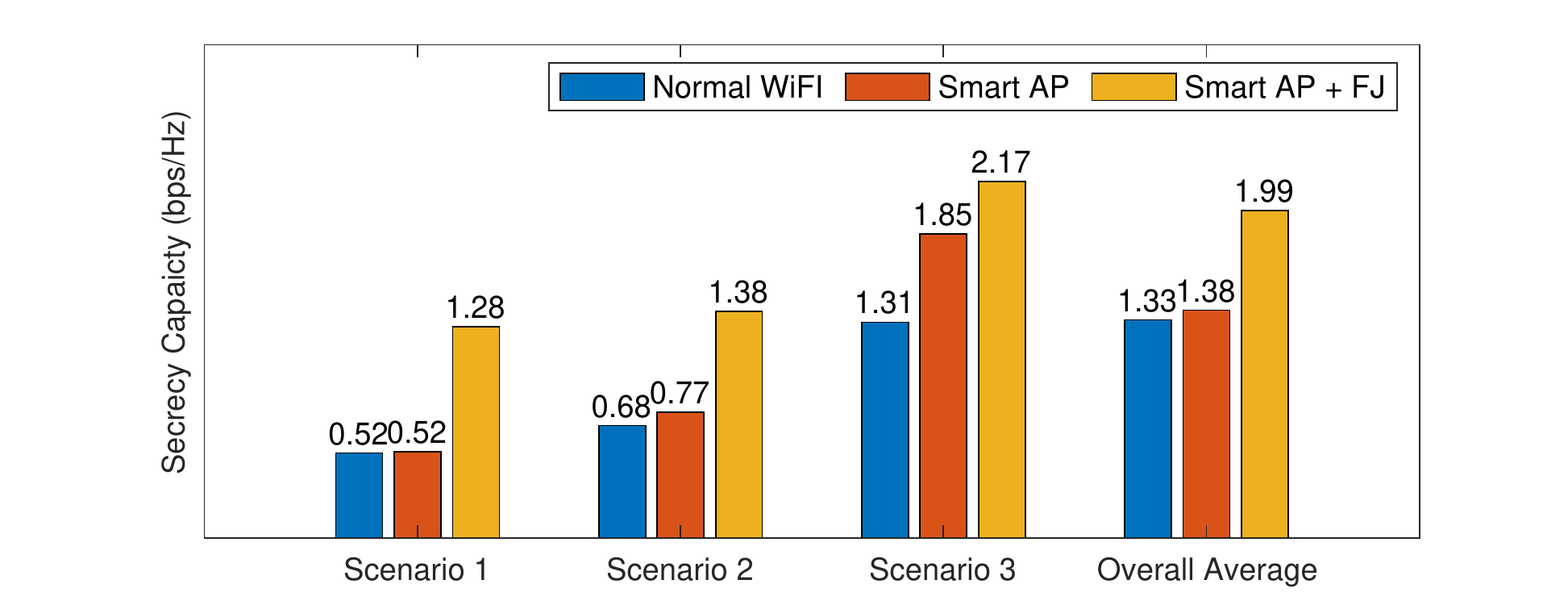}
        \vspace{-6mm}
         \caption{Secrecy Capacity}
         \label{fig:barSec}
         \end{subfigure} \\
         
    \begin{subfigure}[b]{0.5\textwidth}
         \centering
        \includegraphics[trim=55 0 0 0, clip,width=\textwidth]{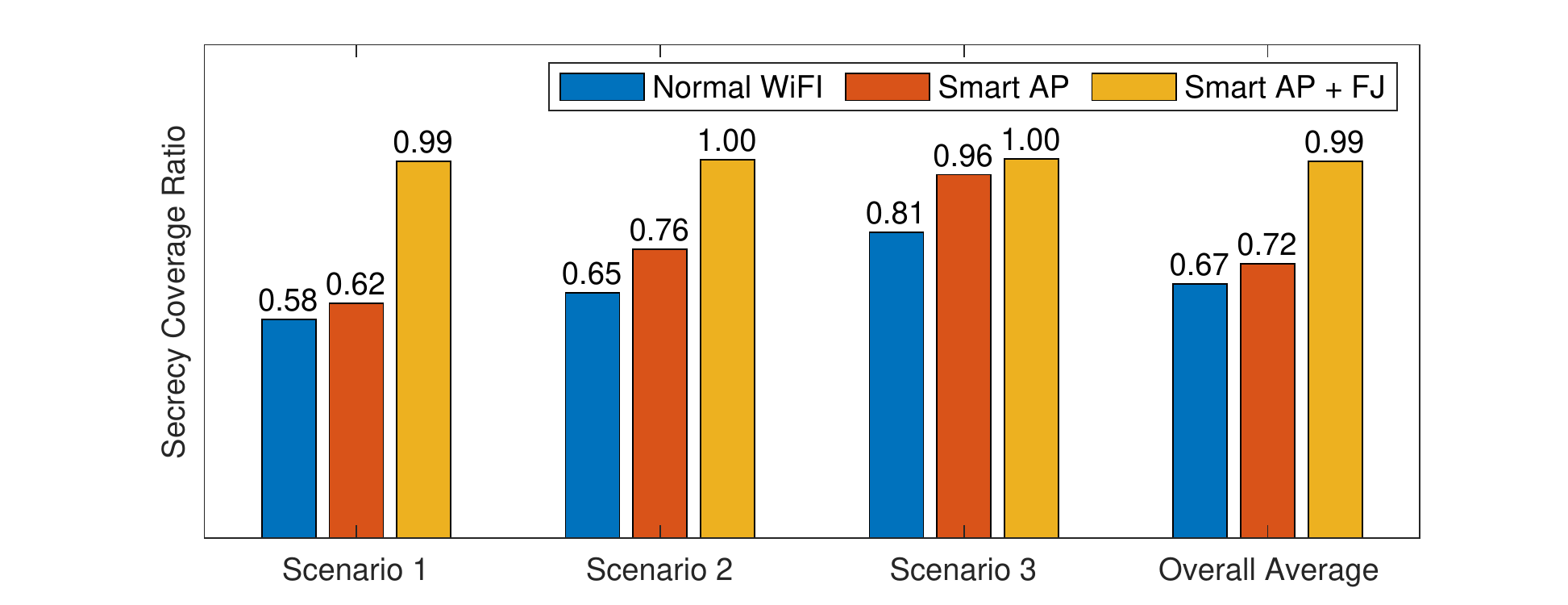}
        \vspace{-6mm}
         \caption{Secrecy Coverage}
         \label{fig:barCover}
         \end{subfigure}
         
    \begin{subfigure}[b]{0.5\textwidth}
         \centering
        \includegraphics[trim=55 0 0 0, clip,width=\textwidth]{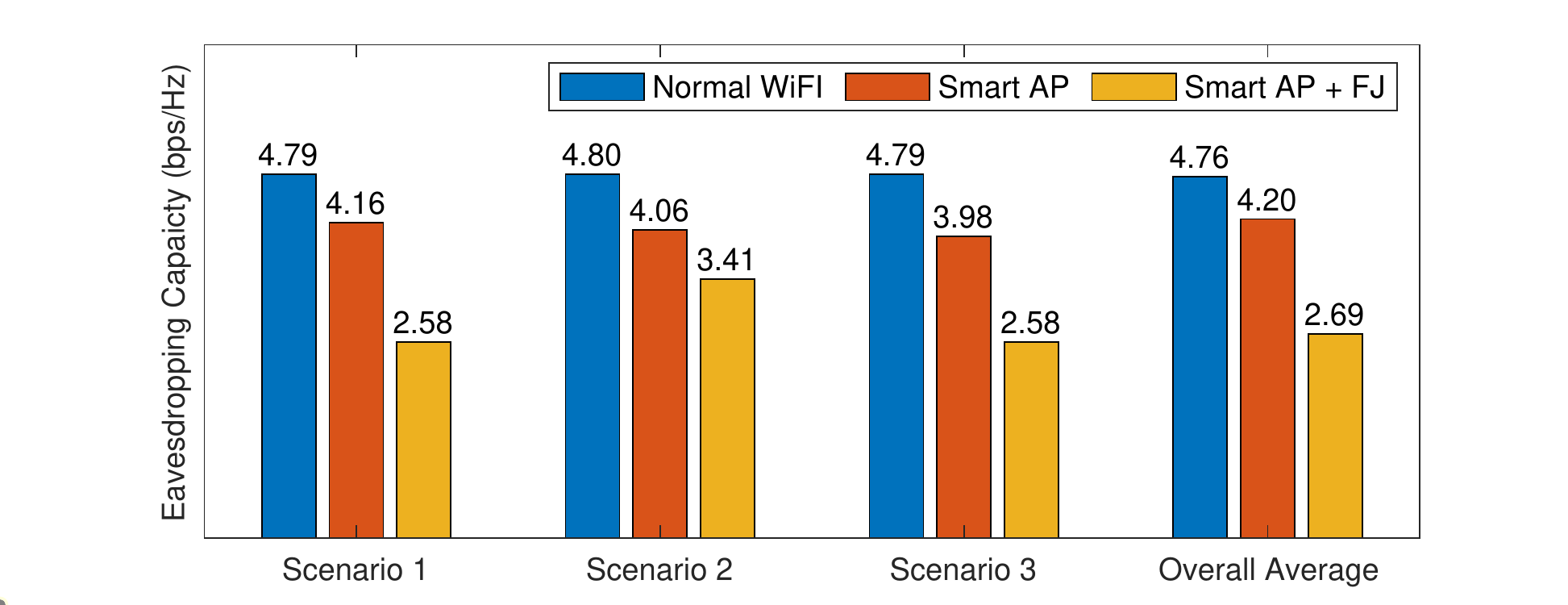}
        \vspace{-6mm}
         \caption{Eavesdropper Capacity}
         \label{fig:barEav}
         \end{subfigure}
         
         \caption{The average secrecy capacity for STA$_m$, the average secrecy coverage ratio, and the average eavesdropper's capacity, for the three individual scenarios. The Overall Average shows the results averaged over 10,000 locations for STA$_m$. The color indicates the secrecy capacity optimization algorithm.}
         \label{fig:barMean}
\end{figure}

Using \eqref{eq:avg_sec_cap} and \eqref{eq:avg_eav_cap}, \figurename~\ref{fig:barMean} shows the average secrecy capacity for STA$_m$, the average secrecy coverage ratio, and the average eavesdropper's capacity for the three individual scenarios. The Overall Average shows the results averaged over 10,000 locations for STA$_m$. The color indicates the secrecy capacity optimization algorithm. The results show that the conclusions from   \figurename~\ref{fig:mapFigures1} can be well generalized to all other locations of STA$_m$. It also shows that adding FJ significantly increases the average secrecy capacity while significantly decreasing the average eavesdropper's capacity. But, arguably, the most interesting result is that the area where no secrecy can be achieved has reduced to almost zero. 

\section{Conclusions and Future Work} 
For a scenario with only two APs controlled by a spectrum-programming enhanced SDN controller, we have shown that perfect secrecy and optimized secrecy capacity can be achieved by means of network-enabled PLS, for nearly every location of the passive eavesdropper, by intelligently combining AP selection for the legitimate station and the generation of FJ by the idle AP. This is an important result, as until now, the applicability of PLS in real networks has greatly suffered under the complexity of its implementation and the inability to secure the network for too many locations of an eavesdropper. Our work shows that both limitations can be overcome by looking at the problem at the level of a programmable network. 

In future work, we will optimize the model and the system further by adding more APs, more legitimate stations, and more eavesdroppers. It can be intuitively understood that a larger scale and a higher density of the wireless network provides more opportunities to further optimize AP selection as well as the generation of FJ. We also intend to validate our simulation results with our testbed as described in \cite{hoseini2022ccnc}. There are also many opportunities to improve the algorithms further. For instance, we here optimized the secrecy capacity by AP selection and FJ in a serial manner: first AP selection, then FJ. The next step is to optimize both mechanisms jointly. The model may also extended by the inclusion  of  other  techniques  such  as  multiple antennae, beamforming  and  network  coding. 

Throughout this paper, we have assumed  that the  eavesdropper’s  channel  condition  is  known,  and  therefore its  location.  This  is  particularly  hard  when  the  eavesdropper  is passive.  To mitigate  this,  we  will investigate if we can consider  the  physical  boundaries  of  the  network,  for instance  an  enterprise  building  or  apartment  block. This makes it possible to distinguish  insider  threats  from  external eavesdroppers, and to make  statistical  estimates  for  the likely  locations  of  the  external  eavesdroppers.  We will also investigate  the  use  of  effective  eavesdropper detection  tools  such  as  Ghostbuster \cite{chaman2018ghostbuster},  which  make use  of  the  fact  that  even  passive  receivers  leak  RF signals, which  can  be integrated  into  the  spectral  programming  architecture. 

We  will  also look at  cases  where  upstream  traffic needs  to  be  confidential  too.  The  relevant  eavesdropper’s Shannon  capacity  is  then  between  STA$_e$ and STA$_m$.  This is  outside  the  control  of  the  spectrum  programming architecture,  unless STA$_m$  is  also  a  controlled  entity.  PLS  can then  in  theory  be  achieved  by  moving  the  AP  closer  to  the STA$_m$.  However,  the  STA$_e$ may  still  be  able  to  decode  the signal  unless link-level  measures  are  being  applied in addition.

\section*{Acknowledgment}
This  work  is  supported  by  the  UNSW  Institute  for  Cyber Security  (IFCYBER) and an internal grant from the School of Engineering and IT, UNSW Canberra.  

\bibliographystyle{IEEEtran}
\bibliography{References}

\end{document}